\documentclass[%
nofootinbib,
 amsmath,amssymb,
]{revtex4-1}
\usepackage{epsf}
\usepackage{graphics}
\usepackage{graphicx}
\usepackage{bm}
\usepackage{color}

\def\MSbar{{\rm  \overline{\footnotesize MS\kern-0.05em}\kern0.05em}}
\def\lrD{\stackrel{\leftrightarrow}{D}}

\newcommand{\pslash}{p \!\!\!/}
\newcommand{\nn}{\nonumber}

\newcommand{\msbar}{\overline{\mbox{MS}}}
\newcommand{\Dlr}{\buildrel \leftrightarrow \over D\raise-1pt\hbox{}\!\!}
 \newcommand{\Dl}{\buildrel \leftarrow \over D\raise-1pt\hbox{}}
\newcommand{\Dr}{\buildrel \rightarrow \over D\raise-1pt\hbox{}}
 
\def\lrD{\stackrel{\leftrightarrow}{D}}

\newcommand{\be}{\begin{eqnarray}}
\newcommand{\ee}{\end{eqnarray}}

\def\VEV#1{\langle #1 \rangle}

\begin{document}

\title{Renormalization of quark propagator, vertex functions and twist-2 operators from
twisted-mass lattice QCD at $N_f$=4} 

\author{Beno\^{\i}t Blossier$^a$, Mariane
Brinet$^b$,
Pierre Guichon$^c$,Vincent Mor\'enas$^d$, Olivier
P\`ene$^a$, Jose Rodr\'iguez-Quintero$^e$ and Savvas Zafeiropoulos$^{d,f}$ }

\affiliation{%
$^a$ Laboratoire de Physique Th\'eorique - CNRS et Universit\'e  Paris-Sud XI \\
B\^atiment 210, 91405 Orsay Cedex France\\
$^b$ Laboratoire de Physique Subatomique et de Cosmologie - Universit\'e Grenoble-Alpes, CNRS/IN2P3, 
53, avenue des Martyrs, 38026 Grenoble, France\\
$^c$ CEA-Saclay, IRFU/SPhN, 91191 Gif-sur-Yvette Fance\\
$^d$ Laboratoire de Physique Corpusculaire - Universit\'e Blaise Pascal,  CNRS/IN2P3, 63000 Aubi\`ere Cedex France\\
$^e$ Dpto. F\'isica Aplicada - Fac. Ciencias Experimentales,\\
Universidad de Huelva 21071 Huelva Spain\\
$^f$Institut f\"ur Theoretische Physik, Goethe-Universit\"at Frankfurt,\\
Max-von-Laue-Str.~1, 60438 Frankfurt am Main, Germany\\
}%

\date{\today}


\collaboration{ETM Collaboration}
\begin{figure}[htpb]
\begin{center}
\includegraphics[width=40mm]{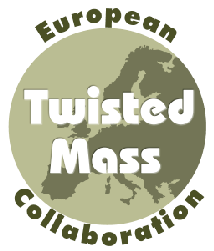}
\end{center}
\end{figure}

\preprint{LPT-Orsay 13-??}
\preprint{UHU-FT/13-??}
\preprint{LPSC????}


\begin{abstract}
We present a precise non-perturbative determination of the renormalization constants in the mass independent RI'-MOM scheme. 
The lattice implementation uses the Iwasaki gauge action and four degenerate dynamical twisted mass fermions. The gauge configurations are provided by the ETM
Collaboration. Renormalization constants for scalar, pseudo-scalar, vector and axial operators, as well as the quark propagator renormalization,
are computed at three different values of the lattice spacing, two volumes and several twisted mass parameters. The method we developed allows for
a precise cross-check of the running, thanks to the particular proper treatment of hypercubic artifacts. Results for the twist-2 operator $O_{44}$
are also presented. 
\end{abstract}

\maketitle



\tableofcontents
\newpage
\section{Introduction}

Lattice QCD (LQCD) has proven to be a very powerful approach to study QCD and has become a precision technique for the ab initio computation of many QCD observables. 
In particular, the possibility to perform a rigorous non-perturbative renormalization is an essential feature of lattice calculations. QCD discretization on a space-time lattice provides indeed a well defined regularization of the theory, by introducing the lattice spacing as a natural cut-off. However, any comparison with physical results requires a precise control of the continuum limit. Renormalization allows, from bare quantities computed at finite lattice spacing, to obtain meaningful physical observables with the accuracy sought (typically of the percent level). Controlling as much as possible all statistical and systematic
effects in the determination of the renormalization constants is crucial since the accuracy of the renormalization procedure directly affects 
the precision of the computed observables. For instance, the calculation of nucleon matrix elements, which remains an open challenge, 
involves a careful estimate of the corresponding renormalization constants, essential to compare lattice results to values deduced from experiments. 
A proper comparison of these matrix elements with experimental values represents both a challenge and an opportunity for lattice QCD. 

The goal of this work is to present the computation of renormalization constants (RCs) for local and twist-2 fermionic bilinear operators using twisted mass
fermion configurations with four dynamical quarks in the sea. 

We use a modified version of the regularization invariant RI-MOM scheme known as RI'-MOM \cite{Martinelli}.
The renormalization conditions of an operator are imposed non-perturbatively on conveniently defined amputated projected Green functions computed
between off-shell quark states, evaluated at a given momentum and in a fixed gauge. This scheme is mass-independent and renormalization constants are defined at zero quark mass. To carry out this renormalization study,  the European
Twisted Mass Collaboration (ETMC) has performed dedicated $N_f=4$ simulations 
with four degenerate light quark masses. RCs for $N_f=2+1+1$ ensembles are evaluated by extrapolating to the chiral limit the RCs computed with the $N_f=4$ ensembles.

By using the lattice formalism, one is obliged to break some symmetries which are only recovered in the continuum limit,
among which is the continuum rotation symmetry. In discrete Euclidean space, the $O(4)$ rotation symmetry
is broken down to $H(4)$ or $H(3)$ hypercubic symmetry depending on whether the lattice setup is the same
on spatial and temporal directions. As a consequence, there are lattice artifacts which are only invariant under $H(4)$ but not under $O(4)$. This is particularly an issue for the computation of quantities
like the renormalization constants since the associated statistical errors are often quite small, and the
uncertainties from lattice artifacts become visible, thus deserving a careful treatment. A popular solution
is to use the "democratic cuts" to select data points with relatively small $H(4)$ lattice artifacts.
Another approach, which is usually called the $H(4)$-extrapolation \cite{H4,Boucaud:2003dx,Feli}, is to include the lattice artifacts explicitly in the data analysis. This approach, applied in the present work, allows one to use a much wider range of data points and to extract information from the lattice simulations more efficiently. 


A particularly interesting point of the $H(4)$-extrapolation procedure applied to the determination of renormalization constants, 
is to allow a precise study of their running. This key advantage provides 
the possibility to compare the evolution of the RCs obtained on the lattice with perturbative formulae and to perform an interesting estimate 
of the non-perturbative contributions.

This paper is organized as follows. After a brief description of the lattice set-up in section \ref{section:latset}, basic RI'-MOM formulae and
our notation are defined in section \ref{section:RIMOM}. The analysis procedure is explained in section \ref{section:analysis}, where
Goldstone pole subtraction, $m_{PCAC}$ average and hypercubic corrections are detailed. A precise study of the running 
is presented in section \ref{section:running}, both for local and twist-2 operators, with a special focus on lattice artifacts and higher order corrections. A comment is in order at this point. In the jargon of the renormalization community there is an abuse of the meaning of the word "local". All operators considered in this analysis such as densities, currents as well as twist-2 operators are of course local from the field theoretical viewpoint. However, it has prevailed within the renormalization community to refer to local operators in particular for the densities and the currents.
Chiral extrapolations are performed in section \ref{section:Xfit} and section \ref{MSbar} presents in detail the way we convert our results to the $\msbar$ scheme. 
In the penultimate section \ref{section:systematics}, we estimate the systematic errors and the final section contains our conclusions.
It is noteworthy that our methods allow for the extraction of the $\langle A^2\rangle$, the Landau gauge dimension-2 gluon condensate~\cite{Boucaud:2001st} that has rich phenomenological implications~\cite{A^2}.

\section{Lattice Set-up}\label{section:latset}

The results presented here are based on the gauge field configurations generated by the ETMC using the Iwasaki gauge action and the twisted mass fermionic action. 
Since the RI'-MOM is a mass-independent scheme, where the renormalization conditions are imposed on the chiral limit, the ETMC has generated dedicated $N_f=4$ ensembles with four light
degenerate quarks \cite{Palao_2011, Italiani}, which would eventually allow for a more trustworthy chiral extrapolation. This is the reason we employ these configurations in our analysis, since the physical configurations with 2 light degenerate $u$ and $d$ quarks and two heavier non degenerate $s$ and $c$ quark would introduce an extra source of rather uncontrolled systematic uncertainty. Of course the results obtained with our configurations are intended to renormalize bare matrix elements which are computed with the physical configurations. To achieve automatic $\mathcal{O}(a)$ improvement, the twisted mass action is usually tuned
to maximal twist, by tuning $m_{PCAC}$ quark mass to zero. However, in the case of four degenerate quarks, reaching the maximal twist 
is far from being a trivial task and an alternative option has been chosen. Ensembles are simulated in pairs, with opposite
values of $m_{PCAC}$, and $\mathcal{O}(a)$ artifacts are removed by averaging the quantities obtained from these 
two ensembles.  Previous studies have indeed shown the feasibility of this approach \cite{Dimopoulos_2010, Palao_2011, Frezzossi_2003}. 
We refer to Ref.~\cite{Palao_2011, Italiani} for more explicit details and for the original discussion of the computational setup. However, for the sake of completion we will remind the reader all the essential aspects of the algorithmic details as well as the parameter tuning of
the ensemble generation. 
More specifically, the full action reads
\be
    \mathcal{S}^{(N_f = 4)} = \mathcal{S}_{g}[U] + \mathcal{S}_{TM}^{\rm sea}[\chi_f^{\rm sea}, U] + \mathcal{S}_{TM}^{\rm val} [\chi_f,\phi_f,U] ~ , 
    \label{completeAction}
\ee
where by $\chi$ we denote the usual "fermionic" quarks both in the sea and the valence sector and by $\phi$ the "bosonic" quarks used in the partial quenching.
The fermionic action for the sea quarks takes the form
\be
   S_{tm}^{\rm sea} = a^4 \sum_{x, f}  \bar\chi_f^{\rm sea} \Big{[}\gamma \cdot \tilde\nabla + W_{cr}  +
                                  (m_{0 f}^{\rm sea} - m_{cr}) + ir_f^{\rm sea} \mu_f^{\rm sea} \gamma_5 \Big{]}\chi_f^{\rm sea} ~ , \\
    \label{actionSea}
\ee
with $\gamma \cdot \tilde\nabla = \frac{\gamma_\mu}{2} (\nabla_\mu + \nabla_\mu^*)$, $W_{cr} = -\frac{a}{2} \nabla_\mu^*\nabla_\mu + m_{cr}$ and $r_d^{\rm sea} = -r_u^{\rm sea}$, $r_c^{\rm sea} = -r_s^{\rm sea}$. 
For the twisted mass parameter the following choice has been made
\be
    \mu_u^{\rm sea} = \mu_d^{\rm sea} = \mu_s^{\rm sea} = \mu_c^{\rm sea} \equiv \mu^{\rm sea} ~ .
    \label{muSea} 
\ee
Accordingly, the fermionic action for the valence quarks takes the form 
\be
    S^{\rm val} = a^4 \sum_{x, f} \bar\chi_f^{\rm val} \Big{[} \gamma \cdot \tilde\nabla -  \frac{a}{2} \nabla_\mu^*\nabla_\mu + m_{0, f}^{\rm val} +  ir_f^{\rm val} \mu_f^{\rm val} \gamma_5 \Big{]} \chi_f^{\rm val} ~ . 
    \label{actionValence}
\ee
The parameters $r_f^{\rm val/sea}$ take the values $\pm 1$, while the twisted masses $a\mu_f^{\rm val,sea}$ are assumed to be non-negative.
Two volumes, three values of the lattice spacing, and several values of the twisted mass have been considered in the analysis. The run parameters are summarized in Table \ref{tab1} and, for illustrative purposes, Table \ref{tab1b} shows the pion masses directly obtained from the appropriate ratio of two-point correlators computed for the same ensembles here analyzed. 

\begin{table}[htdp]
\caption{{\it{ $N_f=4$ ensembles used in our analysis. \label{tab1}}}}
\begin{small}
\begin{center}
\begin{tabular}{|l|c|c|c|c|}\hline
ensemble  & $\kappa$  & $am_{PCAC}$& $a\mu$ ($a\mu_{sea}$ in bold)  & confs $\#$ \\\hline
\hline
\multicolumn{5}{|c|}{$\beta=2.10$ - $32^3.64$}\tabularnewline
\hline
3p &  0.156017 &+0.00559(14) &0.0025, {\bf{0.0046}}, 0.0090, 0.0152, 0.0201, 0.0249, 0.0297  & 250   \\\hline
3m &  0.156209 &-0.00585(08) &0.0025, {\bf{0.0046}}, 0.0090, 0.0152, 0.0201, 0.0249, 0.0297  & 250   \\\hline
4p &  0.155983 & +0.00685(12)&0.0039, {\bf{0.0064}}, 0.0112, 0.0184, 0.0240, 0.0295  & 210   \\\hline
4m &  0.156250 & -0.00682(13)&0.0039, {\bf{0.0064}}, 0.0112, 0.0184, 0.0240, 0.0295  & 210   \\\hline
5p &  0.155949 &+0.00823(08)  &0.0048,  {\bf{0.0078}}, 0.0119, 0.0190, 0.0242, 0.0293 & 220   \\\hline
5m &  0.156291 &-0.00821(11) & 0.0048,  {\bf{0.0078}}, 0.0119, 0.0190, 0.0242, 0.0293 & 220   \\\hline
\hline
\multicolumn{5}{|c|}{$\beta=1.95$ - $24^3.48$}\tabularnewline
\hline
2p &  0.160826 &+0.01906(24) &{\bf{0.0085}}, 0.0150, 0.0203, 0.0252, 0.0298  & 290   \\\hline
2m &  0.161229 &-0.02091(16)  &{\bf{0.0085}}, 0.0150, 0.0203, 0.0252, 0.0298  & 290   \\\hline
3p &  0.160826 & +0.01632(21)&0.0060, 0.0085, 0.0120, 0.0150, {\bf{0.0180}}, 0.0203, 0.0252, 0.0298  & 310   \\\hline
3m &  0.161229 & -0.01602(20)&0.0060, 0.0085, 0.0120, 0.0150, {\bf{0.0180}}, 0.0203, 0.0252, 0.0298  & 310   \\\hline
8p &  0.160524 &+0.03634(14) &{\bf{0.0020}}, 0.0085, 0.0150, 0.0203, 0.0252, 0.0298& 310   \\\hline
8m &  0.161585 &-0.03627(11)   & {\bf{0.0020}}, 0.0085, 0.0150, 0.0203, 0.0252, 0.0298 & 310   \\\hline
\hline
\multicolumn{5}{|c|}{$\beta=1.90$ - $24^3.48$}\tabularnewline
\hline
1p &  0.162876 &+0.0275(04) &0.0060, {\bf{0.0080}}, 0.0120, 0.0170, 0.0210, 0.0260  &   450  \\\hline
1m&  0.163206 &-0.0273(02) &0.0060, {\bf{0.0080}}, 0.0120, 0.0170, 0.0210, 0.0260  &   450  \\\hline
4p &  0.162689 &+0.0398(01) &0.0060, {\bf{0.0080}}, 0.0120, 0.0170, 0.0210, 0.0260  &   370  \\\hline
4m&  0.163476 &-0.0390(01) &0.0060, {\bf{0.0080}}, 0.0120, 0.0170, 0.0210, 0.0260  &   370  \\\hline
\end{tabular}
\end{center}
\end{small}
\end{table}%

\begin{table}[htdp]
\caption{{\it{ $N_f=4$ Pseudoscalar masses for the ensembles used in our analysis. The values correspond to 
the same twisted masses, $a\mu$,  of Table.~\ref{tab1}} and appear displayed in the same order\label{tab1b}}.}
\begin{small}
\begin{center}
\begin{tabular}{|l|c|}\hline
ensemble  & $a m_\pi$ \\\hline
\hline
\multicolumn{2}{|c|}{$\beta=2.10$ - $32^3.64$}\tabularnewline
\hline
3p &    
0.1268(13),  {\bf{0.1417(11)}}, 0.1755(8), 0.2198(7), 0.2516(7), 0.2805(7), 0.3077(7) \\\hline
3m & 
0.1122(22),  {\bf{0.1332(16)}}, 0.1710(13), 0.2161(10), 0.2476(9), 0.2761(8), 0.3028(7)   \\\hline
4p & 
0.1462(12),  {\bf{0.1623(11)}}, 0.1958(9), 0.2431(8), 0.2768(8), 0.3078(8)   \\\hline
4m &  
0.1304(16),  {\bf{0.1532(11)}}, 0.1910(9), 0.2396(8), 0.2734(7), 0.3042(7)   \\\hline
5p &  
0.1606(14),  {\bf{0.1786(12)}}, 0.2051(11), 0.2496(9), 0.2803(9), 0.3088(8)   \\\hline
5m &   
0.1424(13),  {\bf{0.1666(10)}}, 0.1967(8), 0.2435(8), 0.2746(7), 0.3032(7)   \\\hline
\hline
\multicolumn{2}{|c|}{$\beta=1.95$ - $24^3.48$}\tabularnewline
\hline
2p &  
 {\bf{0.2795(16)}}, 0.3107(12), 0.3391(10), 0.3659(9), 0.3909(8)   \\\hline
2m &  
 {\bf{0.2630(16)}}, 0.2987(12), 0.3287(11), 0.3562(10), 0.3815(9)   \\\hline
3p &  
0.2566(18), 0.2678(16),0.2851(14), 0.3016(13),  {\bf{0.3188(12)}}, 0.3318(12), 0.3599(10), 0.3858(9)   \\\hline
3m & 
0.1933(12), 0.2196(12), 0.2503(10), 0.2738(9),  {\bf{0.2956(9)}}, 0.3113(9), 0.3434(8), 0.3717(7)   \\\hline
8p &   
  {\bf{0.3645(16)}}, 0.3723(14), 0.3881(11), 0.4052(10), 0.4232(9), 0.4413(8)  \\\hline
8m &   
  {\bf{0.3253(16)}}, 0.3364(14), 0.3569(12), 0.3776(11), 0.3985(10), 0.4188(10)   \\\hline
\hline
\multicolumn{2}{|c|}{$\beta=1.90$ - $24^3.48$}\tabularnewline
\hline
1p &  
0.3308(13),  {\bf{0.3362(12)}}, 0.3494(12), 0.3698(11),0.3886(10), 0.4132(9)  \\\hline
1m &  
0.2912(12),  {\bf{0.2998(11)}}, 0.3183(10), 0.3431(9), 0.3640(8), 0.3902(8)  \\\hline
4p &   
0.4032(13), {\bf{ 0.4061(12)}}, 0.4138(12), 0.4284(11), 0.4412(11), 0.4598(10)  \\\hline
4m &    
0.3492(12),  {\bf{0.3541(11)}}, 0.3653(10), 0.3840(10), 0.4002(9), 0.4218(8)  \\\hline
\end{tabular}
\end{center}
\end{small}
\end{table}%

The lattice spacing values are respectively $a=0.062(2)$ fm for $\beta=2.10$, $a=0.078(3)$ fm for $\beta=1.95$  and $a=0.086(4)$ fm for $\beta=1.90$~\cite{Palao_2011, Italiani}. 
Strictly speaking setting the scale in the $N_f=4$ theory is not plainly well defined. Physical units such as fm or MeV are by definition linked to a physical theory. In practice,  as no world containing four light degenerate quarks exists and can produce an experimental observation to compare a lattice result with, the value of $a$ for our set-up, as would happen for any other non-physical one, should result from a convention or assumption. Thus, 
guided by the assumption that a pion built out of 2 valence quarks of mass $\mu_{val}=\mu_{sea}$ living in either the $N_f=4$ or $N_f=2+1+1$ theory acquires the same coupling constant in both cases, the ETMC takes the values of $a$ to be the same for both theories. Corrections to this assumption, which ChPT can in principle account for, should surely exist although they are not expected to be very large. Therefore, as the non-perturbative renormalization of $N_f=2+1+1$ lattice QCD in the massless quark limit needs the scale setting for the $N_f=4$ theory, it is natural to take for the lattice spacing of the latter the one of the $N_f=2+1+1$ physical theory (that one is of course also not truly physical, as there is some minor correction from the bottom sea quark, which is however maybe 3 orders of magnitude smaller than the errors of the computation).

Furthermore, in Ref.~\cite{Boucaud:2013bga} a novel procedure for the lattice scale setting is proposed and claimed to be particularly in order for relative "{\it calibrations}" (ratios of lattice spacings) and for non-physical cases such as $N_f=4$ simulations. The very point of the procedure is the scale setting via the inter-comparison of a purely gluonic quantity, as the strong coupling in Taylor scheme, which is first assumed and "{\it a posteriori}" numerically shown not to depend on the quark masses, at least far away from the flavor thresholds. Applied to the $N_f=4$ theory \cite{Boucaud:2013bga}, the resulting lattice spacings are proven to be  compatible with those obtained for ETMC for $N_f=2+1+1$~\cite{Palao_2011, Italiani}.

The fixing of the Landau gauge is achieved by the iterative minimization of a functional of the $SU(3)$ links with a combination of stochastic overelaxation and Fourier acceleration.

\section{Renormalization constants in the RI'-MOM scheme} \label{section:RIMOM}

In this section we define the notation that we utilize, recall briefly the RI'-MOM scheme \cite{Martinelli} and the explicit formulae that will be used in the computation. We refer the reader to Ref.~\cite{Vladikas_LesHouches} for a complete and pedagogical introduction to the RI'-MOM scheme.
The RI'-MOM scheme is very widely used by many lattice collaborations \cite{Gockeler_1999,Martha_2012,Martha_2010,Blossier1,Martha_2013,Italiani}.

\subsection{Basics of RI'-MOM}

We consider a generic bilinear fermion operator $O_{\Gamma}=\bar{q}_1\Gamma q_2$ where $\Gamma$ 
is any Dirac structure, possibly multiplying a covariant derivative operator, 
and $q_1, q_2$ two fermionic fields. In the case of scalar, pseudo-scalar, vector and axial renormalization constants, $\Gamma=1,\gamma_5,\gamma_{\mu},\gamma_5\gamma_{\mu}$
respectively. To avoid contributions from disconnected diagrams, we focus
mainly on the non-singlet quark operators, unless stated differently. The corresponding renormalized operator is defined as 
$O_R=Z_O O_{\Gamma}$  $Z_O$ is found in the RI' variant of the RI-MOM scheme by imposing at a scale $\mu$ large enough
(typically $\mu>>\Lambda_{QCD}$), that the 
amputated Green function in a fixed gauge (the Landau gauge in our case), equals its tree value, i.e. requiring that
\begin{eqnarray}
Z_O(\mu)Z^{-1}_q(\mu)\Gamma_O(p)|_{p^2=\mu^2}=1,
\end{eqnarray}
$Z_q$ is the fermion field renormalization constant, determined through 
\begin{eqnarray}
Z_q(\mu^2=p^2)=-\frac{i}{12p^2}\mbox{Tr}[S^{-1}_{bare}(p)\pslash],
\end{eqnarray}
where $S_{bare}(p)$ is the bare quark propagator. At finite lattice spacing, the four-vector $p$ can be either the
continuum momentum, or the lattice momentum $ap_{\mu}=\sin(ap_{\mu})$. Both definitions differ only by $\mathcal{O}(a^2)$
terms. Since RCs obtained using the continuum momentum already exhibit lattice spacing artifacts at tree-level, the
lattice momentum definition is favored.  $\Gamma_O$ is defined in terms of the amputated Green function, or vertex, $\Lambda_O$, by
\begin{eqnarray}
\Gamma_O(p)={1\over12}\mbox{Tr}\left(\Lambda_O(p)\hat{P}_O\right) \label{GammaO}\;\;
\mbox{  with}\;\;\;\Lambda_O(p)=S^{-1}_{q1}(p)G_O(p)S^{-1}_{q2}(p), \label{LambdaO}
\end{eqnarray}
where $\widehat{P}_O$ is a suitable projector (see section below) and the Green function is defined in coordinate space by
\begin{eqnarray}
G_O(x,y)&=&<q_1(x)\,O_{\Gamma}\,\bar{q}_2(y)>.
\end{eqnarray}
On the lattice and in Fourier space, the Green function becomes
\begin{eqnarray}\label{FFT}
G_O(p)=\int d^4x d^4y\;e^{-ip(x-y)}\,G_O(x,y)&=&{1\over N} \sum_i S_{q_1;i}(p|z)\Gamma \gamma_5 S_{q_1;i}^{\dagger}(p|z)\,\gamma_5,
\end{eqnarray}
where the sum runs over $N$ configurations, $S_i(p|z)=\int d^4 y\,S_i(y,z)e^{-ipy}$ and $\Lambda_O(p)$ in definition Eq.~(\ref{GammaO}) reads
\begin{eqnarray}
\Lambda_O(p)=S^{-1}_{q_1}(p)G_O(p)\gamma_5S^{-1\dagger}_{q_1}(p)\gamma_5. 
\end{eqnarray}
It involves only one type of quark propagators since we have taken into account the properties of the twisted-mass formulation 
relating mass degenerate quarks.

\bigskip

We will study in particular in this work the twist-2 operator $O_{44}$.
Twist-2 operators are of particular importance since they provide the leading contribution to the Operator Product Expansion (OPE) analysis of the deep inelastic scattering and this particular one is associated with the $\langle x\rangle_q =\int_0^1 dx \;x \left(q (x) +\bar q(x)\right)$ of the hadrons \cite{Negele}, where $x$ is the momentum fraction carried by the quark and $q(x)$ the associated longitudinal distribution.

For a general twist-2 operator, $O_{\Gamma}(z,z')=\bar{q}_1(z)\Gamma(z,z') q_2(z')$, we take $\Gamma=\Gamma_{\mu}\lrD_{\nu}$ with $\lrD_{\nu}={1\over 2}({\nabla_{\nu}+\nabla^{*}_{\nu}})$, where $\nabla$ and $\nabla^*$ are respectively the gauge covariant forward and backward derivatives, defined by
\begin{eqnarray}
\nabla_{\nu}(x,y)&=&a^{-1}(\delta_{y,x+\nu}\;U_{\nu}(x)-\delta_{x,y}),\\
\nabla^{*}_{\nu}(x,y)&=&a^{-1}(\delta_{x,y}-\delta_{y,x-\nu}\;U^{\dagger}_{\nu}(x-\nu)),
\end{eqnarray}
where $U$ are the gauge links. Inserting these definitions into the Green function and performing the Wick contractions lead to 
\begin{eqnarray}
G_O(x,y)&=& - {1\over 2} \left\{ S_{q_1}(x,z) \gamma_5 S^{\dagger}_{NL}(y,z) \gamma_5 + S_{NL}(x,z) \gamma_5S^{\dagger}_{q_1}(y,z)\gamma_5 \right\},
\end{eqnarray}
where we have defined 
\begin{eqnarray}
S_{NL}(x,z)= S_{q_1}(x,z+\nu)U^{\dagger}_{\nu}(z)\Gamma_{\mu}  - S_{q_1}(x,z-\nu) U_{\nu}(z-\nu) \Gamma_{\mu}.
\end{eqnarray}
This "non-local propagator" $S_{NL}$, combining the neighboring propagators, is the solution of a Dirac equation with a modified source 
\begin{eqnarray}
\sum_{y}D(x,y) S_{NL}(y,z)= \delta_{x,z+\nu}U^{\dagger}_{\nu}(z)\Gamma_{\mu}-\delta_{x,z-\nu}U_{\nu}(z-\nu) \Gamma_{\mu},
\end{eqnarray}
where $D(x,y)$ is the Dirac operator. Using these "non-local" propagators, from which we can construct all $\Gamma$ structures,
we decrease the number of propagators to be computed, from nine to five (1 with a "local" source and 4 -- one in each direction -- with a  "non-local" one). The advantage of our method is thus the reduced computational cost, since with only 5 inversions per configuration , we are able to extract all local and twist-2 renormalization constants for all momenta. 

Finally, the Green function in momentum space becomes
\begin{eqnarray}
G_O(p)&=& - {1\over 2} .{1\over N} \sum_i  \left\{ S_i^{q_1}(p|z)\gamma_5 S^{q_1\dagger}_{i,NL}(p|z)\gamma_5+S^{q_1}_{i,NL}(p|z)\gamma_5S^{q_1\dagger}_i(p|z)\gamma_5 \right\}.
\end{eqnarray}

\subsection{Projectors}

For scalar and pseudo-scalar operators, the projector $\widehat{P}_O$ in (\ref{GammaO}) is simply $\gamma_0$ and $\gamma_5\gamma_0$ respectively. 
For vector and axial vertex functions however, "naive projectors" $\gamma_{\mu}$ and $\gamma_5\gamma_{\mu}$ do not project vertices 
onto different Lorentz structures. Indeed, the vertex function decomposes over the Dirac structures as
\begin{eqnarray}
\Gamma^V_{\mu}&=&\Sigma^V_1\,\gamma_{\mu}+\Sigma^V_2\,{p_{\mu}\over p^2}\,\pslash ,\\
\Gamma^A_{\mu}&=&\Sigma^A_1\,\gamma_5\,\gamma_{\mu}+\Sigma^A_2\,\gamma_5\,{p_{\mu}\over p^2}\,\pslash , 
\end{eqnarray}
with $\Sigma^V_{1,2}$ and $\Sigma^A_{1,2}$ being scalars multiplied by $3\times3$ identity matrices which we omit to simplify
the notation. The correct projectors are actually given by
\begin{eqnarray}\label{Def:ProjCorr}
P^V_{\mu}&=&\gamma_{\mu}-{p_{\mu}\over p^2}\,\pslash ,  \nonumber\\
P^A_{\mu}&=&\gamma_5\,\gamma_{\mu}-\gamma_5\,{p_{\mu}\over p^2}\,\pslash ,
\end{eqnarray}
and the corresponding form factors are then
\begin{eqnarray}
\Sigma^{A/V}_1={1\over 12}{p^2\over p^2-p^2_{\mu}} \mbox{Tr}\left[ P^{A/V}_{\mu}\Gamma^{A/V}_{\mu} \right].
\end{eqnarray}

We have checked that the effect of using or not using these correct projectors has a rather small influence on $Z_V$ and $Z_A$. However, since 
statistical errors will turn out to be also small, we will use in what follows the correct projectors of Eq.~(\ref{Def:ProjCorr}).

\bigskip
In a similar way, twist-2 operators should be projected such that Lorentz structures are decoupled. Following the convention used in Ref.~\cite{Gracey},
we define a general symmetric and traceless twist-2 operator as
\begin{eqnarray}
O_{\mu\nu}=\gamma_{\mu}D_{\nu}+\gamma_{\nu}D_{\mu}-{1\over 2}\delta_{\mu\nu}\gamma_{\rho}D^{\rho}.
\end{eqnarray}
The corresponding Green function can be decomposed as
\begin{eqnarray}
G_{O}=-{1\over2}\Sigma_1(p)\;\left(\gamma_{\mu}p_{\nu}+\gamma_{\nu}p_{\mu}-{1\over 2}\delta_{\mu\nu}\pslash\right) 
- \Sigma_2(p)  \pslash\left( p_{\mu}p_{\nu} -{1\over 4} p^2 \delta_{\mu\nu}\right).
\end{eqnarray}
To project out the first form factor $\Sigma_1(p)$ we use (correcting minor typos in Ref.~\cite{Gracey})
\begin{eqnarray}
P_O=-p^2 \frac{\left( {p_{\mu}p_{\nu}\pslash\over p^2}-{(\gamma_{\mu}p_{\nu}+\gamma_{\nu}p_{\mu})\over 2}\right)}
{\left( 4p^2_{\mu}p^2_{\nu}-p^2(p^2_{\mu}+p^2_{\nu})-2p_{\mu}p_{\nu}p^2\delta_{\mu\nu}\right)}.
\end{eqnarray}
In particular, in the case of $O_{44}$ operator, we obtain
\begin{eqnarray}
P_O=p^2 \frac{\left( {p^2_{4}\pslash\over p^2}-\gamma_{4}p_{4}\right)}
{4p_4^2\left(\vec{p}^{\;2}\right)}.
\end{eqnarray}

%
%

\section{Analysis procedure}\label{section:analysis} 

For each value of the sea quark mass, two sets of gauge fields are produced, with opposite $m_{PCAC}$ values, corresponding to opposite values of the
angle $\theta$, the latter being defined as the complementary to the twisted angle, see Ref.~\cite{Palao_2011}. The first step consists in removing the
Goldstone pole from vertex functions, for each ensemble, and in performing the valence quark mass extrapolation. The $\theta$ average is then done, before 
correcting for $H(4)$ artifacts. 
The different steps of the analysis are detailed below using a given set of ensembles, namely $3p/3m$ ensembles on a $32^3\times 64$ lattice. 
The results presented in the next two sections concern charged currents and densities: $O=\bar{u}\Gamma d$ or $O=\bar{d} \Gamma u$.
All plots in this section represent average of jackknife bins and statistical errors appear also estimated by the jackknife method. A sensible full-correlated-matrix analysis would require much higher statistics than the one from the available data. This is why we proceed otherwise and apply the jackknife approach for the error analysis also followed in a series of papers~\cite{H4, Boucaud:2001st, Boucaud:2003dx, Blossier1, Blossier_alphaS_1, Blossier:2013ioa}, devoted in the last few years to the study of the non-perturbative running of renormalization constants, and that revealed itself as very useful for dealing with the fits of correlated parameters. Although the meaning for the $\chi^2$/d.o.f. estimates are purely indicative and only useful for comparative purposes, the estimated 
errors from the jackknife bins for the fitted parameters appear to be meaningful.

\subsection{Pion mass and Goldstone pole subtraction}

\begin{figure}[htbp]
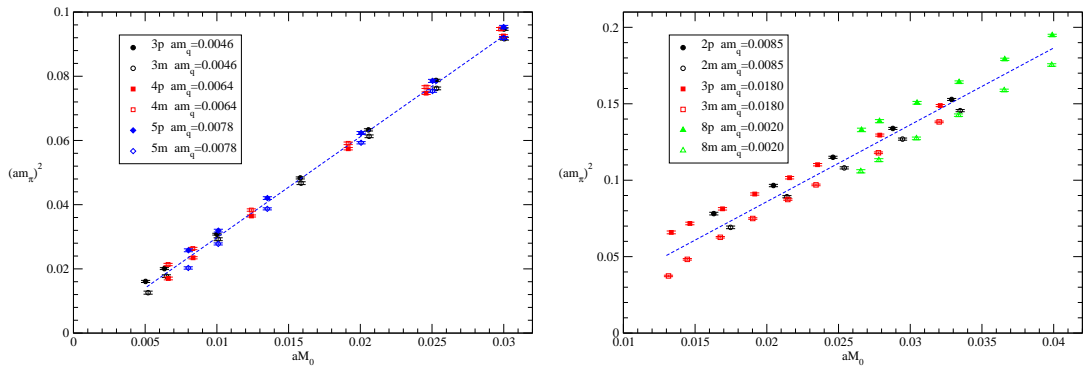
--
\begin{center}
\epsfxsize=7cm\epsffile{figs/pion_mass_beforeaverage_L32T64_9-54_vs_Mrenorm_2.eps}\hspace{0.2cm}
\epsfxsize=7cm\epsffile{figs/pion_mass_beforeaverage_L24T48_6-42_vs_Mrenorm.eps}\hspace{0.2cm}
\caption{{\it {Pion mass for each $\beta=2.10$ (LHS) and $\beta=1.95$ (RHS) ensemble, before $\theta$ average. The $x$-axis is the renormalized quark mass $M_{0}
=\sqrt{(Z_A m_{PCAC})^2+m_q^2}$ and the $y$-axis is the pion mass squared, both in lattice units.
The difference between m/p ensembles illustrates the consequence of non maximal twist and ${\cal{O}}(a)$ effects. The result of the straight line fit using pion mass values computed after $\theta$ average is shown in dashed blue curve.}}}
\label{mass2}
\end{center}
\end{figure}

For each ensemble, the first step of the RCs analysis consists in subtracting the Goldstone pole contribution on vertex functions. This requires to compute the pion mass for each configurations set. 
Pion masses are determined before performing the $\theta$ average. The results are illustrated in Fig.~\ref{mass2}, showing the pion mass as a function of the renormalized quark mass 
$M_{0}=\sqrt{(Z_A m_{PCAC})^2+m_q^2}$ where an estimate of $Z_A=0.78(0.73)$ has been taken for $\beta=2.10$  from Ref.~\cite{Palao_2011} (and $\beta=1.95$ from Ref.~\cite{Petros}).

After the jackknife bins of vertex functions (or $Z_q$, for the quark wavefunction renormalization constant) have been computed, an average over $H(3)$ invariants is performed. 
Then the Goldstone pole subtraction is done, bin by bin and before the $\theta$
average. The pole is taken as the charged pion mass squared at non maximal twist
\begin{eqnarray}\label{Xfit}
\Gamma(p^2,\mu_{sea})=A(p^2)+B(p^2)m^2_{\pi}+{C(p^2)\over m^2_{\pi}},
\end{eqnarray}
where $C(p^2)$ is the  non perturbative Goldstone pole contribution \cite{Cudell}. 
The value of the subtracted vertex functions $\bar{u}\Gamma d$ and $\bar{d}\Gamma u$ extrapolated to zero mass 
are very similar. This
justifies the average over non-singlet operators that will be performed later in the analysis.  
As we only consider charged vertices, there is to our knowledge no coupling
to any neutral Goldstone pole. The extrapolation to the chiral limit of RC's, 
instead of being performed in term of the quark masses, is done in function
of the pion mass. 
There is strong corroborative evidence, based on the Vacuum Saturation Approximation in support of the fact that the charged pion mass is much less affected by $\mathcal{O}(a^2)$ effects than the neutral pion mass \cite{cutoffEffects}. 
The important statement is that the dominant contamination of the charged vertices in question comes from the charged pion. This is also shown explicitly in \cite{Vladikas_LesHouches}.
Of course, there are neutral pions in the sea, but whatever contribution they give must be subdominant.
 
\begin{figure}[htbp]
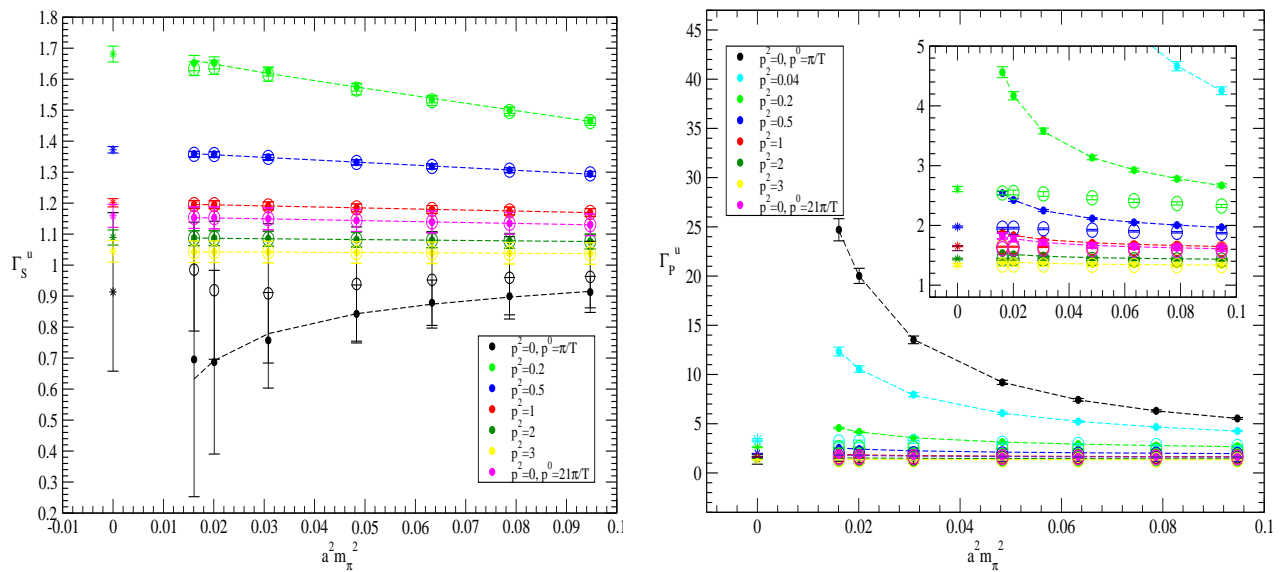

\begin{center}
\epsfxsize=8.2cm\epsfysize=7.5cm\epsffile{figs/GammaS_3p_u_vs_mpi2.eps}\hspace{0.3cm}
\epsfxsize=8.2cm\epsfysize=7.5cm\epsffile{figs/GammaP_3p_u_vs_mpi2.eps}\caption{{\it {
$\bar{d}\Gamma u$
 scalar (LHS) and pseudo-scalar (RHS) vertex functions versus pion mass squared (in lattice units) for ensemble $3p$ ($\beta=2.10$) for several values of $a^2\vec{p\;}^2$ ($ap^0={\pi\over T}$ for all curves except the magenta one, for which $ap^0={21 \pi\over T}$). }}}
\label{GammaSP_1}
\end{center}
\end{figure}

Only the pseudo-scalar vertex is expected to exhibit a Goldstone pole. We have however also inspected other vertices to check possible contamination
or lattice artifacts. 
On Figure \ref{GammaSP_1} for the $3p$ ensemble, the scalar vertex functions for the $u$ quark are plotted (filled symbols) versus the pion mass squared and compared with the subtracted values (empty circles). The extrapolated value is also indicated (star symbol). The difference between subtracted and non-subtracted values lies at the fourth digit for all valence quark masses for $a^2\vec{p\;}^2 \ge \sim 0.5$, and it is not visible on the plot. 
For lower $a^2\vec{p\;}^2$ values however, the subtraction effect is visible. Since no Goldstone pole contamination is expected for $Z_S$
\cite{Vladikas_LesHouches}, those effects are likely to be lattice artifacts.  A similar conclusion holds for the quark renormalization constant and for the vector current. 
The axial vector current also has a coupling to the pion and this could be potentially a source of a problem but since this coupling is proportional to the momentum transfer of the process, which actually vanishes for our kinematical setup, it poses no problem either \cite{Pagels,Gattringer}.

We stress that, since the low values of $a^2\vec{p\;}^2$ will be excluded from 
the fit range considered when compensating for $H(4)$ artifacts (see next section), these lattice effects will not influence the final results.

Contrary to $\Gamma_S$, the pseudo-scalar vertex function shows, as expected, a strong pion mass dependence. Vertex functions for ensembles
$3p$ are displayed in the RHS of Fig.~\ref{GammaSP_1}, with the same legend conventions as for the plots of the scalar vertex. 
The Goldstone pole appears clearly and is thus subtracted according to Eq.~(\ref{Xfit}).  

\vspace{0.9cm}

\begin{figure}[htbp]
\vspace{0.9cm}
\begin{center}
\epsfxsize=8cm\epsfysize=7.5cm\epsffile{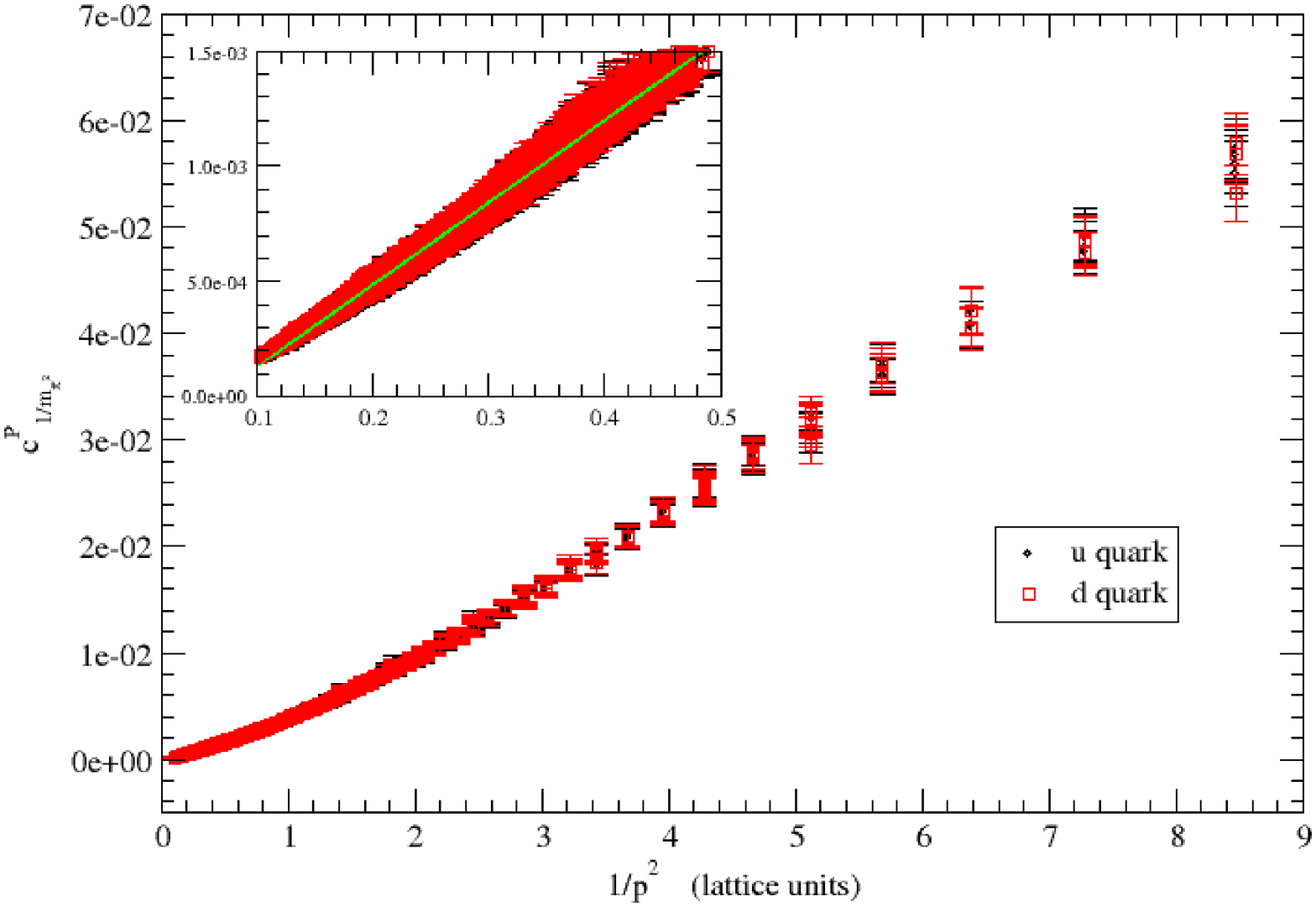}\hspace{0.5cm} 
\epsfxsize=8cm\epsfysize=7.5cm\epsffile{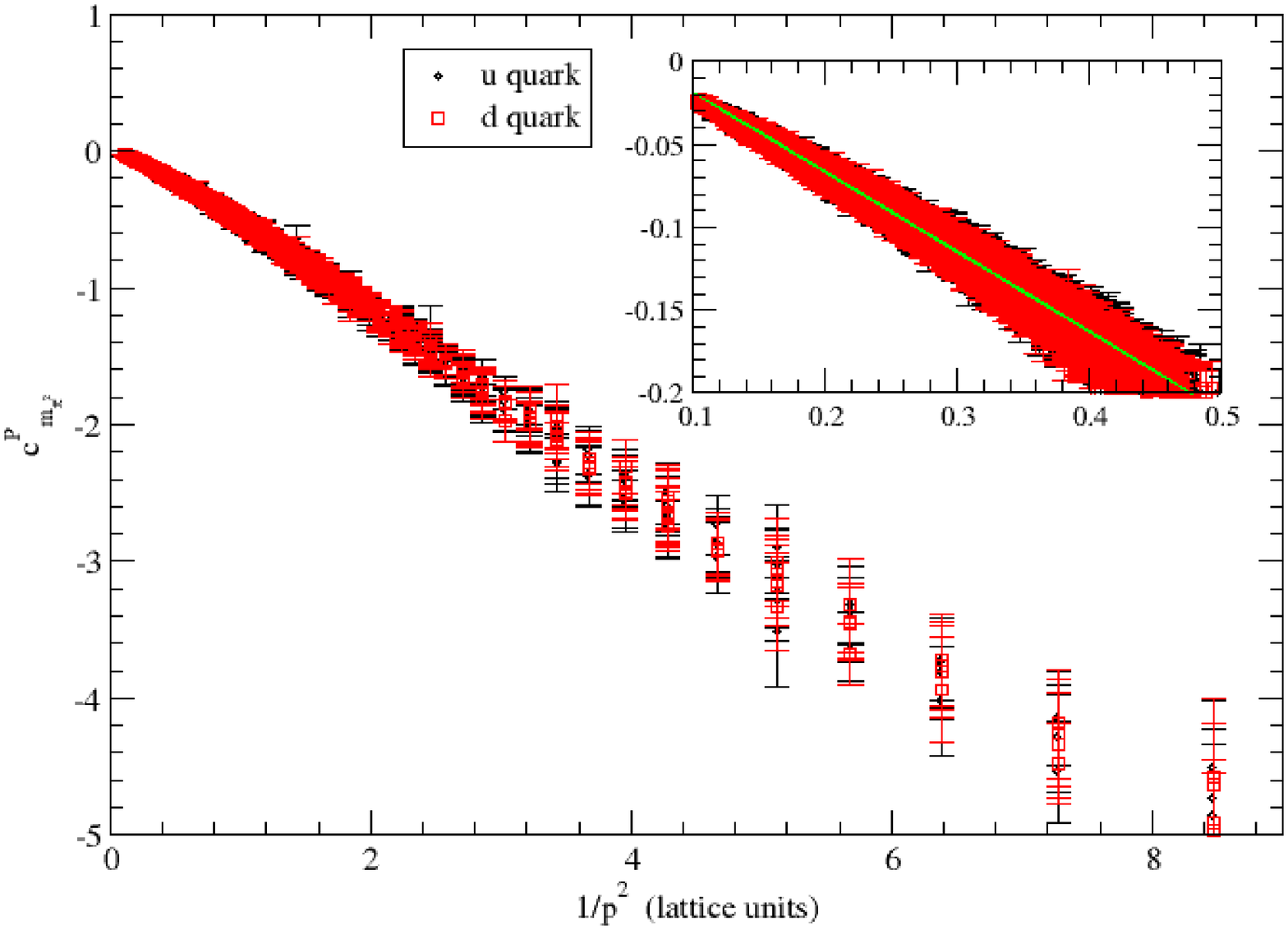} 
\caption{{\it {Coefficient of the $1/m^2_{\pi}$ term (l.h.s.) and of the $m^2_{\pi}$ term (r.h.s.) in the chiral fit (in (\ref{Xfit}), $C(p^2)$ and $B(p^2)$ resp.) 
as a function of ${1/p^2}$ in lattice units, for ensemble $3p$ ($\beta=2.10$). The green line serves as eye guidance mainly and represents a linear fit at large $p^2$.}}}
\label{coeff1}
\end{center}
\end{figure}

The $p^2$ dependence of the chiral extrapolation coefficients are displayed in Fig.~\ref{coeff1}. 
The $1/m^2_{\pi}$ coefficient of the chiral extrapolation is, as expected, varying as $1/p^2$ for pseudo-scalar vertex, at
large $p^2$. Over the large range of $a^2p^2$ values considered (typically for $a^2p^2>0.1$), this coefficient varies as
$c_1/p^2+(c_2/p^2)^2$. This is consistent with the expectation that the Goldstone pole can only appear in power
suppressed non-perturbative contributions. For other vertices, it is globally compatible with zero. 

\subsection{$m_{PCAC}$ average and Hypercubic corrections}
 
\begin{figure}[htbp]
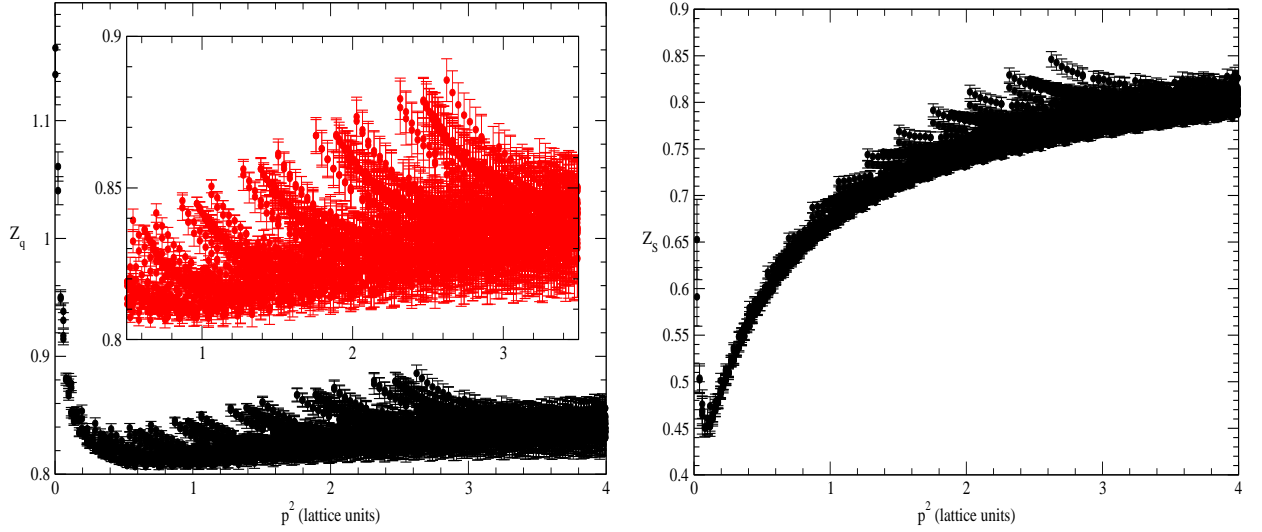

\vspace{0.9cm}
\begin{center}
\epsfxsize=8cm\epsfysize=7.cm\epsffile{figs/Zq_mPCACaveraged_struct00_afterH3_beforeH4.eps}\hspace{0.3cm}
\epsfxsize=8cm\epsfysize=7.cm\epsffile{figs/ZS_mPCACaveraged_struct00_afterH3_beforeH4.eps} 
\caption{{\it { Quark renormalization constant (LHS) and scalar renormalization constant (RHS.) as a function of $p^{2}$ in lattice units. Both exhibit the typical 
"fishbone" structure induced by the breaking of the $O(4)$ rotational symmetry of the Euclidian space-time by the lattice discretization, into the hypercubic group $H(4)$.}}}
\label{after_mPCAC1}
\end{center}
\end{figure}

 The $m_{PCAC}$ average is performed on the vertex function jackknife bins. Since they differ only by (small) lattice artifacts, non-singlet operators ($\bar{u}Od$ and $\bar{d}Ou$) are also averaged at this stage, with an average weighted by jackknife errors. The scalar (RHS) and quark wave function (LHS) renormalization constants are represented for the representative pair ensembles $3m/3p$ (see Table \ref{tab1}) in Fig.~\ref{after_mPCAC1} 
 as a function of $p^{2}=p_{\mu}p^{\mu}$, in lattice units. 
$Z_q$ exhibits the usual strong half "fishbone" structure, typical of hypercubic artifacts, while other renormalization constants are also affected, although to a lesser extent. 

The three vector components $\gamma_{\mu}$ (and similarly for the three axial ones $\gamma_{\mu}\gamma_5$) being very similar (not shown on these Figures) and degenerate in the continuum limit, they are also averaged before hypercubic artifacts are removed. 

\begin{figure}[htbp]
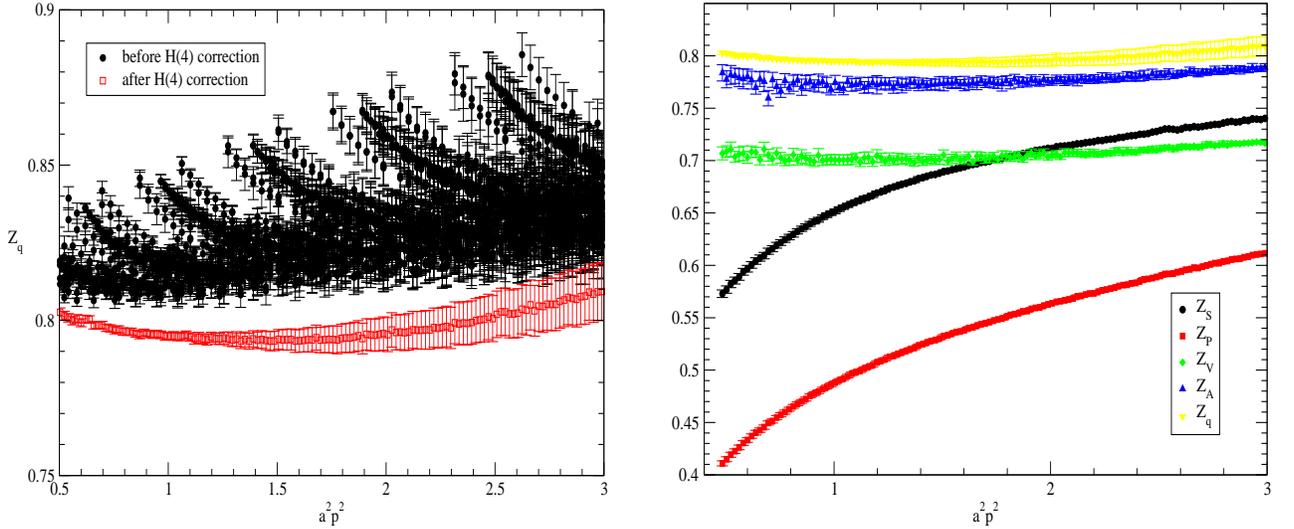

\vspace{0.9cm}
\begin{center}
\epsfxsize=8cm\epsfysize=7cm\epsffile{figs/Zq_mPCACaveraged_struct00.eps} 
\hspace{0.6cm}
\epsfxsize=8cm\epsfysize=7cm\epsffile{figs/AllRCs_mPCACaveraged_Charged_CorrProj_H4onRC_OWF_3mp_L32T64.eps} 
\caption{{\it {LHS: Effect of hypercubic corrections on quark renormalization constant, as a function of $a^2p^{2}$, for ensemble $3mp$ ($\beta=2.10$, $\mu_{sea}=0.0046$, $32^3\times64$ lattice). RHS: renormalization constants as a function of $a^2p^{2}$, after removing $H(4)$ artifacts, still for ensemble $3mp$ ($\beta=2.10$, $\mu_{sea}=0.0046$, $32^3\times64$ lattice).}}}
\label{after_H4_1}
\end{center}
\end{figure}

The next step consists in correcting one of the two types of $\mathcal{O}(a^2)$ artifacts, namely the hypercubic artifacts which respect the $H(4)$ 
symmetry group but not the $O(4)$ one (the second type, i.e. $\mathcal{O}(a^2p^2)$ artifacts, respecting the continuum $O(4)$ rotation symmetry, 
will be treated non-perturbatively
by introducing corrections in the running of the RCs, see section \ref{section:running}). A very powerful method has been developed \cite{H4,Feli}, which does not rely on  
the selection of a small subset of momenta, thus keeping the maximum amount of information. 
A by-product of this procedure is the fact that, unlike other methods, 
it allows to test the running of the renormalization constants. This method has already been extensively and fruitfully exploited in studying the QCD running coupling \cite{Blossier:2013ioa,Blossier:2013te,Blossier_alphaS_1,Blossier_alphaS_2,Blossier:2010ky,Boucaud:2008gn} or the gauge fields propagators~\cite{H4,Boucaud:2005xn,Ayala:2012pb}, while its applications to renormalization have been presented in detail in \cite{Blossier1,Boucaud:2005rm,Boucaud:2003dx}. It will only be recalled here briefly for the sake of consistency. 
We define the following $H(4)$ invariants
\begin{eqnarray}
p^{[2]}=\sum_{\mu=1}^{4}p_{\mu}^2  , \qquad p^{[4]}=\sum_{\mu=1}^{4}p_{\mu}^4, \qquad p^{[6]}=\sum_{\mu=1}^{4}p_{\mu}^6 , \qquad p^{[8]}=\sum_{\mu=1}^{4}p_{\mu}^8,
\end{eqnarray}
and denote the quantity $Z(ap_{\mu})$ (representing any renormalization constant) averaged over the cubic orbits as
$Z_{latt} (a^2 p^2, a^4p^{[4]}, a^6p^{[6]}, ap_4, a^2\Lambda_{QCD}^2)$. We then assume (and will check) that $Z_{latt}$ can be Taylor-expanded
around $p^{[4]}=0$ up to values of $\epsilon=a^2p^{[4]}/p^2$ significantly larger than 1 as
\begin{eqnarray}
Z_{latt} (a^2 p^2, a^4p^{[4]}, a^6p^{[6]}, ap_4, a^2\Lambda_{QCD}^2)=Z_{hyp_{}corrected} (a^2 p^2, ap_4, a^2\Lambda_{QCD}^2)+
R(a^2p^2,a^2\Lambda_{QCD}^2)\frac{a^2p^{[4]}}{p^2}+ \ldots,
\end{eqnarray}
with 
\begin{eqnarray}
R(a^2p^2,a^2\Lambda_{QCD})=\frac{dZ_{latt} (a^2 p^2, a^4p^{[4]}, a^6p^{[6]}, ap_4, a^2\Lambda_{QCD}^2)}{d\epsilon}|_{\epsilon=0},
\end{eqnarray}
$R(a^2p^2,a^2\Lambda_{QCD})$ being reasonably well approximated by $R(a^2p^2,a^2\Lambda_{QCD}^2)=c_{a2p4}+c_{a4p4}a^2p^2$. We use the one window fit technique, described in detail in \cite{Blossier1}.
The fitting range in $a^2p^{2}$ is chosen to be $[0.5-3]$. For values of $a^2p^{2}$ larger than $\approx 3$, some orbits start are indeed missing because
the Fourier transform Eq.~(\ref{FFT}) has been limited to $[-{\pi\over 2},+{\pi\over 2}]$. The effect of the hypercubic correction is clearly
seen for $Z_q$, on the LHS of Fig.~\ref{after_H4_1}. The same procedure is applied to all renormalization constants, leading to the results of 
 Fig.~\ref{after_H4_1} (RHS), which summarizes the results of all local RCs as a function of $a^2p^{2}$. 
The hypercubic correction can be applied either at the vertex level, or, as we did, directly on the renormalization constants. We checked
that there is no significant difference between these two choices. 

\begin{figure}[htbp]
\vspace{0.9cm}
\begin{center}
\epsfxsize=8cm\epsfysize=7cm\epsffile{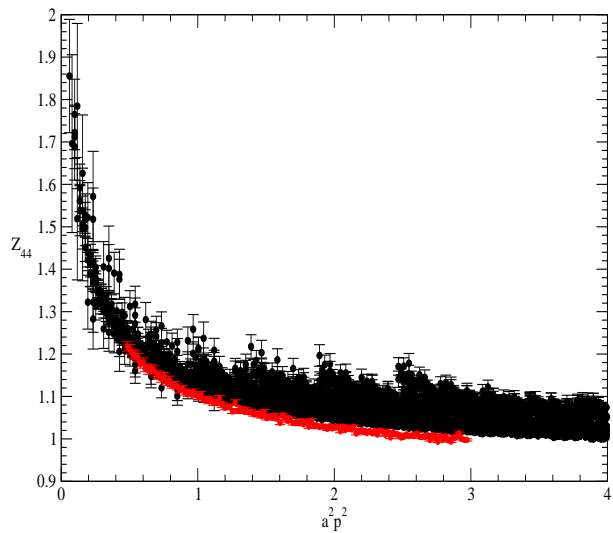} 
\caption{{\it {Effect of hypercubic corrections on $Z_{44}$, as a function of $a^2p^{2}$, for ensemble $3mp$ ($\beta=2.10$, $\mu_{sea}=0.0046$, $32^3\times 64$ lattice).}}}
\label{Z44_1}
\end{center}
\end{figure}

The analysis for twist-2 operators is similar (except for the valence quark mass extrapolation) and applied here to $O_{44}$.
Fig.~\ref{Z44_1} displays the renormalization constant for $Z_{44}$ before and after the hypercubic corrections. 
The "fishbone" structure is about twice more pronounced than for $Z_q$ for instance. 

Finally, we note that there are ultraviolet artifacts which are functions of $a^2p^2$ and are thus insensitive to hypercubic
biases and not corrected by the above-mentioned method. They will be corrected simply by assuming  $a^2p^2$ terms in the final running fit. 

As a summary, our analysis procedure to extract renormalization constants consists in the following steps
\begin{itemize}
\item the $H(3)$ average of the vertex functions over $\vec{p}^{\;2}$ orbits, 
\item the $1/m^2_{\pi}$ term subtraction and valence chiral extrapolation, also done at the level of the vertex function, for each four-momentum, 
\item the $\theta$ average, performed on $Z_q$ or at
the vertex level for other renormalization constants, 
\item the average over non-singlet $\bar{u}Od$ and  $\bar{d}Ou$ operators, weighted by jackknife errors,
\item the average over equivalent $\mu$ ($=1,2,3$) directions for vector and axial operators,  
\item the correction of hypercubic artifacts using an efficient and well-defined procedure.
\end{itemize}

The running of the RCs will be described in details in the next section. 

\section{Running and $\mathcal{O}(a^2p^2)$ artifacts}\label{section:running}

The possibility to check the running of renormalization constants is an important feature of our analysis. This allows to study remaining lattice artifacts 
and non-perturbative contributions and to finally extract reliable values of the RCs. 

\subsection{Renormalization constants for quark and local operators}

We consider the following expression for the running of the quark wave function RC 
\begin{eqnarray}\label{Eq:Zqrun1}
Z_q^{hyp-corr}(a^2p^2)&=& Z^{pert\,RI'}_q(\mu^{2})\,c^{RI'}_{0Z_q}({p^2\over \mu^{2}},\alpha(\mu))\nonumber \\
&\times&\left( 1 +   \frac{ \VEV{A^2}_{\mu^2}}{32p^2}
\frac{c_{2Z_q}^{\bar{MS}}({p^2\over \mu^2},\alpha(\mu))}
{c_{0Z_q}^{RI'}({p^2\over \mu^2},\alpha(\mu))}\frac{c_{2Z_q}^{RI'}({p^2\over \mu^2},\alpha(\mu))}
{c_{2Z_q}^{\bar{MS}}({p^2\over \mu^2},\alpha(\mu))} \right)+  c_{a2p2}\; a^2\,p^2 +  c_{a4p4}\;(a^2p^2)^2 ,
\end{eqnarray}
which was derived in Ref.~\cite{Blossier1} using an OPE analysis. The coefficients $c_{0Z_q}^{RI'}$ and $c_{2Z_q}^{\bar{MS}}$ are known 
from perturbation theory and the running formula contains lattice artifact terms $\propto a^2p^2$ and $\propto (a^2p^2)^2$, not yet removed. These
additional terms are discussed below. We are left with four parameters to determine, namely the value of the RC $Z^{pert\,RI'}_q(\mu^{2})$ at a given renormalization scale $\mu$, the dimension-2 Landau gauge gluon condensate $\VEV{A^2}_{\mu^2}$ and the coefficients $c_{a2p2}$ and $c_{a4p4}$. 
The expression (\ref{Eq:Zqrun1}) includes, apart from the corrections accounting for the not-yet-removed artifacts, the  non-perturbative power correction, $1/p^2$, generated by the non-vanishing gluon condensate~\cite{Boucaud:2001st,A^2}. The same non-perturbative contributions have been previously proven to be mandatory when describing the running of gluon, ghost and quark propagators renormalization constants~\cite{Boucaud:2001st,Boucaud:2008gn,Boucaud:2005xn,Boucaud:2005rm}. We find here, as will be illustrated below, that Eq.~(\ref{Eq:Zqrun1}), nothing less but nothing more, perfectly describes the quark wave function. For the other renormalization constants we also deal with, the same non-vanishing gluon condensate must contribute, {\it via} their corresponding OPE expansion, to generate the same non-perturbative power corrections, unless the Wilson coefficient is proven to be zero. However, even for the latter case, other possible non-perturbative corrections as the pion pole for $Z_P$, hadron contributions or lattice artifacts may also contribute and be hardly disentangled from the OPE ones, with the present level of precision. Then, as will be seen in the following, for renormalization constants other than the quark wave function, we will apply a power correction in lattice units,  $1/p_{\rm latt}^2=1/a^2p^2$, and will not distinguish the different possible sources for its origin. The coefficient of this correction is $a$ dependent and is such that cancels the naive divergence. Unfortunately, due to the fact that we only have three different values of the lattice spacing in the available ensembles we can not perform an accurate fit in order to determine the functional behavior of this coefficient.

We illustrate below the results for $Z_q$ with the ensemble $3mp$, which is representative of the results we obtained with all other ensembles. 
We take  $\Lambda_{QCD}=316$ MeV from Ref.~\cite{Blossier_alphaS_2} and $a_{\beta=2.10}=0.062$ fm from Ref.~\cite{Palao_2011}. The results for local renormalization constants are not sensitive to these values and changing  $a$ and $\Lambda_{QCD}$ over a wide range induces only a change in the local RCs values on the last digit. This however is not the case for twist-2 operators 
 (see section \ref{T2_results}).

\begin{figure}[htbp]
\begin{center}
\vspace{0.9cm}
\epsfxsize=8cm\epsfysize=6.5cm\epsffile{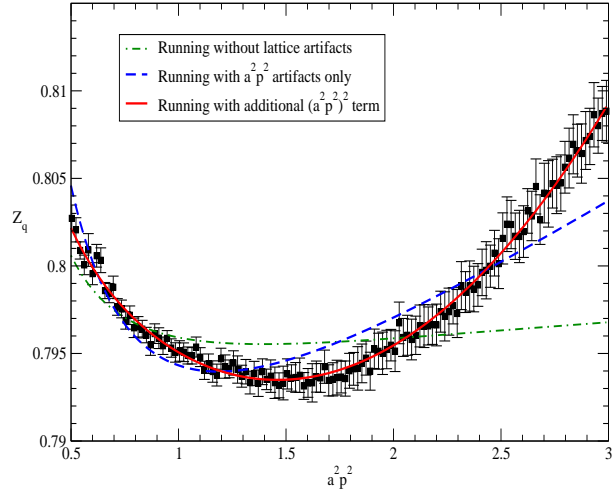}
\caption{{\it {Running of $Z_q$ for ensemble $3mp$ ($\beta=2.10$, $\mu=0.0046$, volume $32^3\times 64$) using different fitting formulae.}}}
\label{Fig:Zqrun3}
\end{center}
\end{figure}

Fig.~\ref{Fig:Zqrun3} displays the running of $Z_q$ for the ensemble $3mp$, fitted by different formulae, depending on whether or not lattice artifacts are included in the running. It can be shown that the standard OPE expression without any artifact correction (dot-dashed green curve) fails completely to describe the data. Adding only an $a^2p^2$ term decreases the $\chi^2/d.o.f.$ down to $2.85$ but the running is still not correctly reproduced. The $\chi^2/d.o.f.$ can be, even more, diminished if the fitting window is restricted down to lower momenta. However, as the $H(4)$-extrapolation has been proved to deal properly with hypercubic artifacts for $Z_q$ up to higher momenta, $a^2p^2 \sim (\pi/2)^2 \sim 2.5$~\cite{Boucaud:2005rm,Blossier1}, we prefer to add only one more parameter to be fitted, including both $a^2p^2$ and $(a^2p^2)^2$ terms accounting for $O(4)$-invariant artifacts, to obtain a good fit 
over the whole range of momenta. The running is then very well described ($\chi^2/d.o.f.=0.26$) and we get at 10 GeV $Z_q(\mu=10 GeV)=0.815(10)$. Errors quoted are at the moment only statistical. Precise estimations of the systematic errors will be performed in section \ref{section:systematics}. 

The same study is performed for scalar and pseudo-scalar RCs. $Z_S$ and $Z_P$ have the same running formula, namely

\begin{eqnarray}
Z_{P/S}(\mu)=Z_{P/S}(\mu_0){c^{RI'MOM}(\mu)\over c^{RI'MOM}(\mu_0)}+ c_{a2p2}\;  a^2\,p^2 +  \frac{c_{p2m1}}{ a^2\,p^2},\label{ZPZSRunning}
\end{eqnarray}
where we have added $1/ (a^2\,p^2)$ and $ a^2\,p^2$ lattice artifact terms . We have \cite{ChetyrkinRetey}
\begin{eqnarray}
c^{RI'MOM}(\mu)&=&x^{\bar{\gamma}_0}  \left\{ 1+(\bar{\gamma}_1-\bar{\beta}_1\bar{\gamma}_0)\; x 
+{1\over 2} \left[  (\bar{\gamma}_1 -\bar{\beta}_1\bar{\gamma}_0)^2  +\bar{\gamma}_2+\bar{\beta}_1^2\bar{\gamma}_0
-\bar{\beta}_1\bar{\gamma}_1-\bar{\beta}_2\bar{\gamma}_0 \right]\; x^2 \right.\nonumber \\
 &+&\left. \left[  {1\over6}  (\bar{\gamma}_1 -\bar{\beta}_1\bar{\gamma}_0)^3+{1\over 2} (\bar{\gamma}_1 -\bar{\beta}_1\bar{\gamma}_0)
(\bar{\gamma}_2+\bar{\beta}_1^2\bar{\gamma}_0-\bar{\beta}_1\bar{\gamma}_1-\bar{\beta}_2\bar{\gamma}_0)  \right. \right.\nonumber\\
&+&\left.\left. {1\over3} (\bar{\gamma}_3-\bar{\beta}_1^3\bar{\gamma}_0+2\bar{\beta}_1\bar{\beta}_2\bar{\gamma}_0-\bar{\beta}_3\bar{\gamma}_0+\bar{\beta}_1^2\bar{\gamma}_1-\bar{\beta}_2\bar{\gamma}_1-\bar{\beta}_1\bar{\gamma}_2)  \right] \; x^3 +{\cal{O}}(x^4)
 \right\},
\end{eqnarray}
where $x=\alpha$, $\bar{\gamma}_i=\gamma_i/\beta_0$  and $\bar{\beta}_i=\beta_i/\beta_0$. $\beta_i$ are the coefficients
of the QCD beta-function and they are given at four-loop in \cite{ChetyrkinRetey}. Their expressions for scalar, pseudo-scalar operators  and quark propagator can be written  \cite{Gockeler_1999,3loops}
\begin{center}
\begin{tabular}{lcllcllcl}
$\beta_0$                   &=&    $11-\frac{2}{3}N_f, \qquad $    &   $\beta_1$  &=&  $102-\frac{38}{3}N_f, \qquad $                 & $\beta_2$&=& $\frac{2857}{2}-\frac{5033}{18}N_f+\frac{325}{54}N_f^2$ ,\\
\end{tabular}
\end{center}
and the anomalous dimensions $\gamma_i$ are given below
\begin{center}
\begin{tabular}{lcllcllcl}
$\gamma^{S/P}_0$  &=&   $-3C_F,\qquad$          &  $\gamma^{S/P}_1$ &=&${1\over 2}\left(-\frac{404}{3}+\frac{40}{9}N_f\right),\qquad$ & $\gamma^{S/P}_2$&=&${1\over2}\left(-2498+\left(\frac{4432}{27}+\frac{320}{3}\zeta(3) \right)N_f+\frac{280}{81}N_f^2\right)$. \\
\end{tabular}
\end{center}

\begin{figure}[htbp]
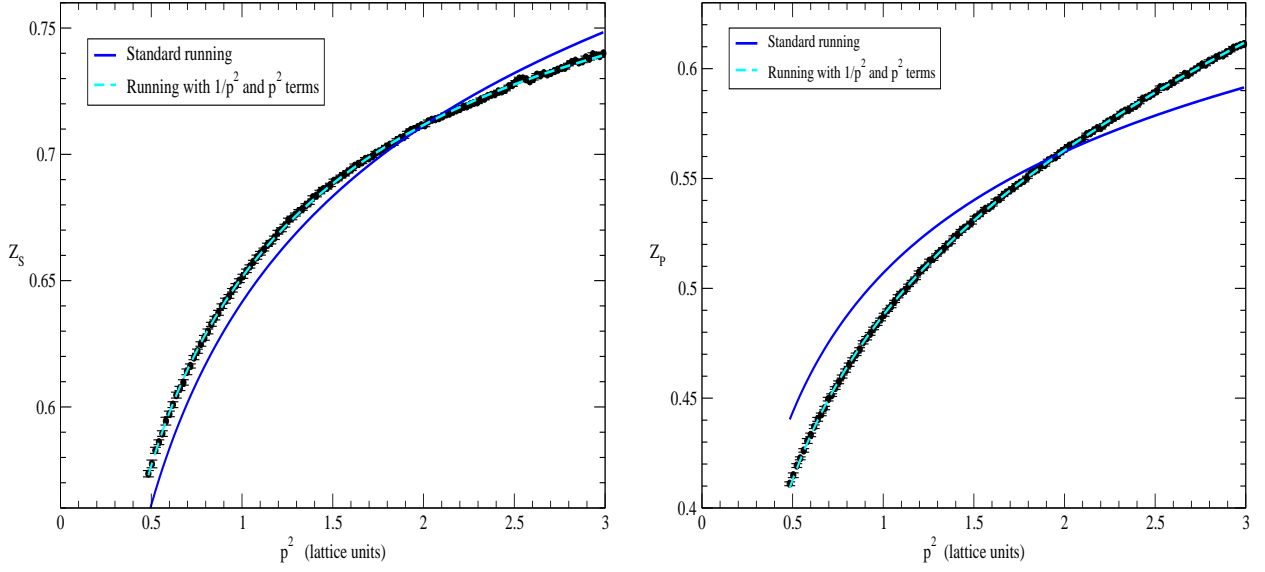

\vspace{0.9cm}
\begin{center}
\epsfxsize=8cm\epsfysize=7.5cm\epsffile{figs/ZS_running_Charged_ProjCorr_H4onRCs_a_AdditionTerms.eps}
\hspace{0.3cm}
\epsfxsize=8cm\epsfysize=7.5cm\epsffile{figs/ZP_running_Charged_ProjCorr_H4onRCs_a_2.eps}
\caption{{\it {LHS: running of $Z_S$ for ensemble $3mp$ ($\beta=2.10$, $\mu=0.0046$, volume $32^3\times 64$). The standard running formula is represented in solid blue line,
the dashed cyan curve includes an $1/p^2$ and an $p^2$ term (in lattice units). This latter fit leads to $Z_S(10 \mbox{ GeV})= 0.869(4)$.
RHS.: Running of $Z_P$ with the standard running expression from  (\ref{ZPZSRunning}) (solid blue curve), and adding an $1/p^2$ and an $p^2$ terms (in lattice units, dashed cyan curve). The modified running gives $Z_P(10 \mbox{ GeV})= 0.623(2)$.  }}}
\label{Fig:ZSZPrun3}
\end{center}
\end{figure}

As for the quark renormalization constant, the standard
running formula, i.e  Eq.~(\ref{ZPZSRunning}) without $1/p^2_{latt}$ and $p^2_{latt}$ terms, fails to describe the running of both $Z_S$ and $Z_P$, 
as illustrated in Fig.~\ref{Fig:ZSZPrun3} (solid blue curves), though to a lesser extent than for $Z_q$.
Additional terms are needed to take into account $\mathcal{O}(a^2p^2)$ artifacts. The evolution of scalar and pseudo-scalar RCs can
be perfectly reproduced with a $1/p^2_{latt}$ and a $p^2_{latt}$ terms added to the standard running by fitting coefficients $c_{a2p2}$ and $c_{p2m1}$, leading to the dashed cyan curves in Fig.~\ref{Fig:ZSZPrun3} 
($\chi^2/d.o.f.=1.14$ and $\chi^2/d.o.f.=0.74$ for respectively $Z_S$ and $Z_P$). The scalar and pseudo-scalar RCs values obtained at 10 GeV for this ensemble are
$Z_S(\mu=10 GeV)=0.869(4)$ and $Z_P(\mu=10 GeV)=0.623(2)$.

\begin{figure}[htbp]
\begin{center}
\epsfxsize=8cm\epsfysize=7.5cm\epsffile{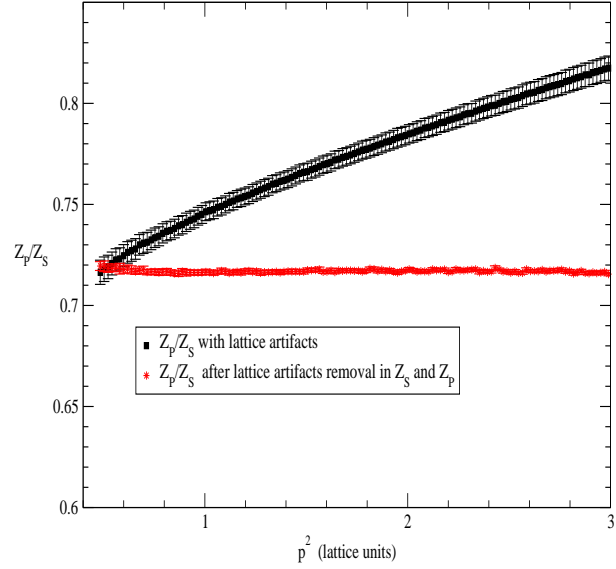}
\caption{{\it {$Z_P/Z_S$ for ensemble $3mp$ ($\beta=2.10$, $\mu=0.0046$, volume $32^3\times 64$). Lattice artifacts have been removed separately from $Z_S$ and $Z_P$. The ratio of these two RCs is compatible with a constant over the whole $p^2$ interval considered and $Z_P/Z_S=0.717(3)$.}}}
\label{Fig:ZSoverZP}
\end{center}
\end{figure}

Scalar and pseudo-scalar RCs having the same anomalous dimension, are expected to have similar running and
their ratio should be constant. If $Z_P/Z_S$ is computed without properly taking into account lattice artifacts, the ratio
varies by more than $20\%$ on the momentum range considered (see Fig.~\ref{Fig:ZSoverZP}, black circles). However, once $\mathcal{O}(a^2p^2)$ artifacts have been separately removed from $Z_S$ and $Z_P$, the ratio becomes compatible with a constant with very good accuracy, over the whole range of $p^2_{latt}$ values (see Fig.~\ref{Fig:ZSoverZP}, red stars). This is an additional indication that
lattice artifacts have been efficiently removed but also that the Goldstone pole has been correctly addressed. 

\begin{figure}[htbp]
\vspace{0.9cm}
\begin{center}
\epsfxsize=8.cm\epsfysize=7cm\epsffile{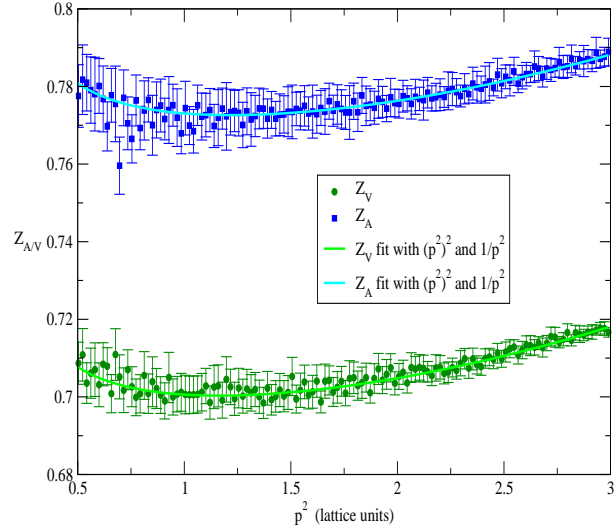}
\caption{{\it {Fits of the residual $p_{latt}^2$ dependence of $Z_V$ and $Z_A$ for ensemble $3mp$ ($\beta=2.10$, $\mu=0.0046$, volume $32^3\times 64$).}}}
\label{Fig:ZAZVrun}
\end{center}
\end{figure}

Axial and vector renormalization constants do not run but it turns out that they exhibit a small $p^2_{latt}$ dependence, which is not
surprising since all other local RCs also show this feature.  
Their variation does not reach more than $4\%$ in total on the momentum range considered, but to extract reliable
values of $Z_V$ and $Z_A$, we remove these artifacts
by fitting this dependence, which turns out to be well described by a combination of $1/p^2_{latt}$ and $(p^2_{latt})^2$ terms. 
The results of the fit are shown in Fig.~\ref{Fig:ZAZVrun}, and lead
to values $Z_V=0.688(5)$ and $Z_A=0.761(4)$.

Table \ref{Tab:summary} summarizes the values obtained for local RCs, for all $N_f=4$ ensembles considered.
\begin{center}
\begin{table}[htbp]
\begin{center}
\begin{tabular}{|l|l|l|l|l|l|l|l|l|}\hline\hline
& \multicolumn{3}{|c|}{$\beta=2.10$ - $32^3\times 64$}  & \multicolumn{3}{|c|}{$\beta=1.95$ - $24^3\times 48$} & \multicolumn{2}{|c|}{$\beta=1.90$ - $24^3\times 48$} \\\hline
 				                                          &	3mp	    	     &	    4mp     	      &	    5mp          & 2mp 	             & 3mp                  & 8mp                  &      1mp	              &       4mp	      \\\hline
$Z^{pert}_q$ ($\mu_R=a^{-1}$)		    &	 0.797(3)     &	  0.785(3)       &	0.787(3)       &  0.763(2)	    &	  0.762(3)       &	0.772(7)        &       0.752(3)          &     0.751(3)         \\\hline\hline
$Z_S$  ($\mu_R=a^{-1}$)	                                &   0.658(3)	     &  0.653(3)        & 0.657(3)         &	0.598(3)	    &  0.603(3)        &  0.601(4)          &    0.582(3)	               &   0.570(3)           \\\hline\hline
$Z_P$  ($\mu_R=a^{-1}$)	                                &   0.472(2)	    &  0.472(2)        & 0.473(2)         &	0.386(2)	    &	0.383(3)       & 0.380(3)           &   0.347(4)	               & 0.343(2)      	     \\\hline\hline
$Z_V$ 				                                & 0.688(5)	    &  0.685(2)        & 0.688(1)         & 0.641(2)  	    & 0.636(2)         & 0.644(7)           &  0.625(3)	                &  0.619(3)     	     \\\hline\hline
$Z_A$ 				                                & 0.761(4)	    &  0.753(2)        & 0.756(1)         & 0.727(2)  	    & 0.725(2)         & 0.733(6)           &  0.721(2)	                &   0.713(2)    	     \\\hline\hline
$Z_P(I=1)/Z_S(I=1)$		                                & 0.717(3)	    &	0.724(3)	     & 0.720(3)         & 0.645(5)	    &  0.634(5)       & 0.632(5)           & 0.597(7)                 &    0.602(5)    \\\hline\hline
\end{tabular}
\end{center}
\caption{{\it {Values of $Z_q$, $Z_S$, $Z_P$, $Z_V$, $Z_A$ and $Z_P/Z_S$ for all $N_f=4$ ensembles analyzed. \label{Tab:summary}}}}
\end{table}%
\end{center}

In order to estimate the uncertainties on the RCs, coming from the lattice spacing determination, we vary $a$ by 5\%, which corresponds to the difference
between the lattice spacing determined using either pion data or nucleon data \cite{Italiani,Martha_2013}. 
We check the influence of this variation on $Z_q$, $Z_S$, $Z_P$ and $Z_P/Z_S$ for a given ensemble, namely $2mp$, $\beta=1.95$. Results are summarized in Table \ref{Tab:avar}.
Scalar and pseudoscalar $Z$ factors are the most sensitive to the lattice spacing, whereas $Z_q$ varies less than one percent and as expected the
ratio $Z_P/Z_S$ is remarkably constant.

\begin{center}
\begin{table}[htbp]
\begin{center}
\begin{tabular}{|l|l|l|l|}\hline\hline
 				      $a$[fm]                      &    0.072         & 	0.078              & 0.084           	      \\\hline
$Z^{pert}_q$ ($\mu_R=$10 GeV)	&     0.785(2)       &	0.786(2)	    	 &  0.788(3)      	     \\\hline
$g^2 \langle A^2 \rangle$[GeV]$^2$	                   &    2.42(10)        &	2.54(10)	       	 &   2.67(10)             \\\hline\hline
$Z_S$  ($\mu_R=$10 GeV)                &     0.840(5)       &	0.859(5)	   	 &  0.879(5)       	     \\\hline\hline
$Z_P$  ($\mu_R=$10 GeV)                &     0.542(3)       &   0.554(3) 	   	 &  0.567(3)              \\\hline\hline
$Z_P(I=1)/Z_S(I=1)$		                   &    0.645(5)     	  &  0.645(5) 	  	 &     0.645(5)    \\\hline\hline
\end{tabular}
\end{center}
\caption{{\it {Dependence of local RCs on a lattice spacing variation, for the ensemble $2mp$.\label{Tab:avar}}}}
\end{table}  
\end{center}

\subsection{Twist-2 operators}\label{T2_results}

The running expression used for $Z_{44}$ is the same than for $Z_S$ (cf. Ref.~\cite{ChetyrkinRetey}, Eq.~ (70))
\begin{eqnarray}
Z_{44}(\mu)=Z_{44}(\mu_0){c^{RI'MOM}(\mu)\over c^{RI'MOM}(\mu_0)}+ c_{a2p2}\; a^2p^2 +  \frac{c_{p2m1}}{a^2p^2}   ,\label{Z44Running}
\end{eqnarray}
and with \cite{ChetyrkinRetey}
\begin{eqnarray}
c^{RI'MOM}(\mu)&=&\mbox{exp}\left\lbrace\int^x \;dx'\frac{\gamma(x')}{\beta(x')}\right\rbrace
=x^{\bar{\gamma}_0}  \left\{ 1+(\bar{\gamma}_1-\bar{\beta}_1\bar{\gamma}_0)\; x\right. \\
&+&\left.{1\over 2} \left[  (\bar{\gamma}_1 -\bar{\beta}_1\bar{\gamma}_0)^2  +\bar{\gamma}_2+\bar{\beta}_1^2\bar{\gamma}_0
-\bar{\beta}_1\bar{\gamma}_1-\bar{\beta}_2\bar{\gamma}_0 \right]\; x^2 +{\cal{O}}(x^3)
 \right\}.
\end{eqnarray}
As for the local RCs, we have added artifacts to the standard running formula, and we fit the coefficients $c_{a2p2}$ and $c_{p2m1}$.
The anomalous dimension for $O_{44}$ is taken from Ref.~\cite{Gracey2} and reminded here for completeness
\begin{eqnarray}
\gamma_{O_{44}}&=&\frac{32}{9}\;a-\frac{4}{243}[378N_f-6005]a^2 \nonumber\\
&+&\frac{8}{6561} [10998N_f^2-6318\zeta(3)N_f-467148N_f-524313\zeta(3)+
3691019]a^3+\mathcal{O}(a^4),
\end{eqnarray}
with $a={g^2\over 16 \pi^2}$. 

\begin{figure}[htbp]
\begin{center}
\epsfxsize=8cm\epsfysize=6.5cm\epsffile{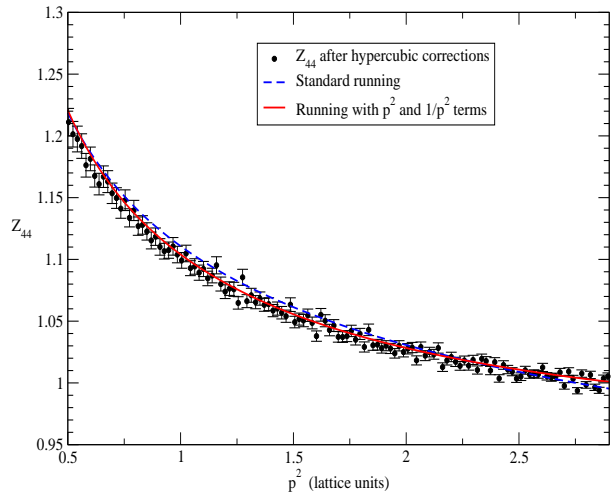}
\caption{{\it {Running of $Z_{44}$ for ensemble $3mp$, $\beta=2.10$, $\mu=0.0046$, $L=32$, $T=64$. The black points are the data after hypercubic artifacts removal. The dashed blue curve is the standard running expression Eq.~(\ref{Z44Running}), and the solid red line includes  $1/(p^2_{latt})$ and $p^2_{latt}$ artifacts.}}}
\label{fig3}
\end{center}
\end{figure}

As can be seen in Fig.~\ref{fig3}, only small lattice artifacts are affecting $Z_{44}$, compared to the case of local RCs. 
When adding $a^2p^2$ and $1/(a^2p^2)$ artifacts to the standard running
expression Eq.~(\ref{Z44Running}), the $\chi^2$ of the fit is decreased and the $Z_{44}$ value changed by $3-5\%$. 

The results are sensitive to the values of the lattice spacing $a$ and of $\Lambda_{QCD}$ at the percent level and the 
uncertainties on both $a$ and $\Lambda_{QCD}$ will be taken into account in the analysis of systematic errors (see section VIII).

%
%
%

\section{Chiral extrapolation and lattice spacing dependence}\label{section:Xfit} 

\begin{figure}[htbp]
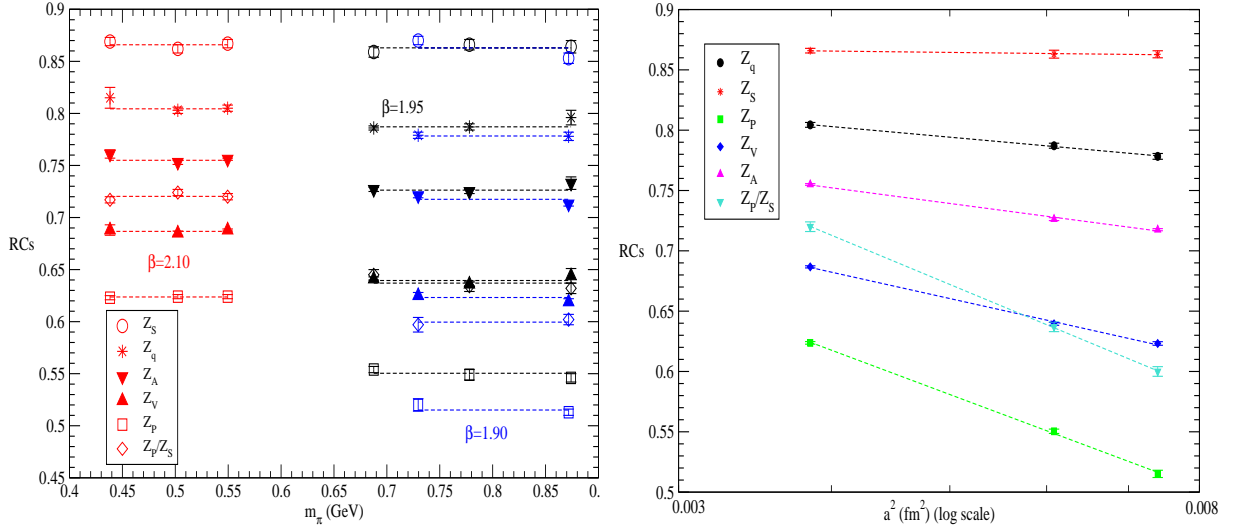

\vspace{0.9cm}
\begin{center}
\epsfxsize=8.cm\epsfysize=7cm\epsffile{figs/AllRCs_vs_Mpi_GeV_ZPoverZS.eps}
\epsfxsize=8.cm\epsfysize=7cm\epsffile{figs/AllRCs_a_dependence_log.eps}
\caption{{\it {LHS: $N_f=4$ local RCs dependence with the pion mass. All RCs are given in the RI'-MOM scheme at 10 GeV. The straight dashed lines are constant fits for each $\beta$ values. The red points
correspond to $\beta=2.10$, the black ones to $\beta=1.95$, and the blue ones to $\beta=1.90$. RHS: local RCs after chiral extrapolation, vs $\log a^2$.
All RCs follow a linear dependence with $\log a^2$ to a very high accuracy. }}}
\label{Fig:Xextrap}
\end{center}
\end{figure}

To get the final values of RCs at each $\beta$ value, we need to perform the chiral extrapolation.
The pion masses are way above the domain of validity of chiral perturbation theory (see Tab.~\ref{tab1b}). 
This could not be avoided for technical reasons, namely
simulation instabilities at low quark masses for $N_f=4$ twisted-mass fermions. 
No theory tells us the expected behavior in terms of $m_\pi$.
Notwithstanding this, some extended prejudice exists about the fact that the $m_\pi$ dependence is small.
Indeed, left-hand side of Fig.~\ref{Fig:Xextrap} displays the pion mass dependence of 
local RCs for the three $\beta$ values under study and, as can be seen, all renormalization constants only depend very weakly on the pion mass, at least within the range where our pion masses lie. 
We thus perform a constant fit to get the chiral limit (dashed lines), in the assumption that the weak dependence we find still works for low pion masses. 
Though supported by the data of the local RCs shown here, this however remains an assumption and the systematic errors from our extrapolation at zero quark mass are poorly controlled. Further results closer to that limit would be helpful to clarify the situation.

Also visible on Fig.~\ref{Fig:Xextrap} is the fact that, if RCs are constant with respect to the pion mass, they are, 
to various extent and with the noticeable exception of $Z_S$, dependent on $\beta$. This is particularly striking on $Z_P, Z_V$ and $Z_A$, and to a lesser extent 
on $Z_q$. To analyze further this variation, we plot in Fig.~\ref{Fig:Xextrap}  (RHS), RCs in the chiral limit versus the lattice spacing squared in logarithmic scale.  
All RCs follow with a very high accuracy a $\log(a^2)$ variation. Although, since RCs are used in practice to renormalize matrix elements computed at a fixed
$\beta$ value, it is not crucial to take this dependence into account in the analysis, it is still interesting to notice that the remaining
lattice spacing dependence is in $\log(a^2)$.

\begin{figure}[htbp]
\vspace{0.9cm}
\begin{center}
\epsfxsize=8.cm\epsfysize=7cm\epsffile{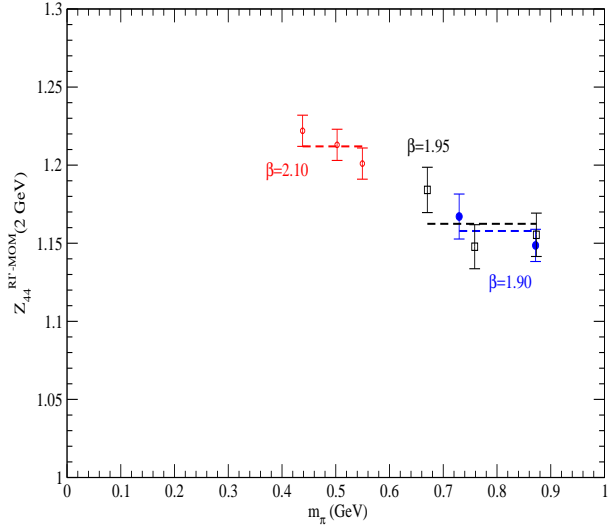}
\caption{{\it {L.h.s.: $N_f=4$ $Z_{44}$ dependence with the pion mass. $Z_{44}$ is given in the RI'-MOM scheme at 2 GeV.
The straight dashed lines are constant fits for each $\beta$ values. The red points
correspond to $\beta=2.10$, the black ones to $\beta=1.95$, and the blue ones to $\beta=1.90$. }}}
\label{Fig:XextrapZ44}
\end{center}
\end{figure}

Unlike local RCs, $Z_{44}$ exhibits a non negligible pion mass dependence, as shown on Figure \ref{Fig:XextrapZ44}. The values of $Z_{44}$ change
by several percents (3-4\%) in the pion mass range considered (440-870 MeV), yet this change is
mainly observed at $\beta=1.95$. For this reason we perform the chiral
extrapolation by a constant fit. 

The  pion mass dependence of $Z_{44}$ will be taken into 
account in evaluation of the systematics.

\section{Conversion to the $\msbar$ scheme and evolution to a reference scale}\label{MSbar}

In order to make the connection with phenomenological calculations and experiments, which almost exclusively refer to the $\msbar$ scheme, we convert
our renormalization factors from RI'-MOM to $\msbar$ using 3-loop perturbative conversion factors obtained from Ref.~\cite{Gracey}. These latter are 
defined as $Z_q^{\msbar}=C^{-1}_q\,Z_q^{RI'-MOM}$ and $Z_{\mathcal{O}}^{\msbar}=C^{-1}_{\mathcal{O}}\,Z_{\mathcal{O}}^{RI'-MOM}$.
In terms of the $\msbar$ coupling constant $\alpha_{\msbar}=\frac{g^2}{16\pi^2}$, and in the Landau gauge, 
these functions read{\protect{\footnote{Setting the covariant gauge parameter $\lambda_{RI'}$ to zero leads to $\lambda_{\msbar}=0$ and since in addition $\alpha_{RI'}=\alpha_{\msbar}+\mathcal{O}(\alpha^5_{\msbar})$ \cite{Gracey}, these conversion functions have the same expression in the Landau gauge whether they are expressed in terms of $\msbar$ or RI'-MOM variables.}}}
\begin{eqnarray}
C_q\;\;\;&=&1+\left[5C_F-(82-24 \zeta(3))\;C_A+28T_FN_f    \right]\;\frac{C_F\alpha^2}{8}\nonumber\\
&+&
\left[   (678024\zeta(3)+22356\zeta(4)-213840\zeta(5)-1274056)\; C_A^2\right. \nonumber\\
&-& (228096\zeta(3)+31104\zeta(4)-103680\zeta(5)-215352 )\;C_AC_F  +31536\;C_F^2 \nn \\
&-&\left.  (89856\zeta(3)-760768)\; C_AT_FN_f+(68256-82944\zeta(3))\;C_FT_FN_f-100480 T_F^2N_f^2  \right]\;\frac{C_F\alpha^3}{5184}+\mathcal{O}(\alpha^4),
\\
C_{S,P}&=&1-4C_F\alpha+[(57-288 \zeta(3))\;C_F  +332T_FN_f+(432\zeta(3)-1285)\;C_A ]\;\frac{C_F\alpha^2}{24} \nn\\
&+&[(-2493504\zeta(3)+155520\zeta(5)+2028348)\;C_AC_F-  (- 3368844\zeta(3)  +466560\zeta(5)+6720046 ) \;C_A^2 \nn\\
&+& (-532224\zeta(3) +186624\zeta(4)+3052384)C_AT_FN_f+(-331776\zeta(3)-186624\zeta(4)+958176)C_FT_FN_f\nn\\
&-&(-451008\zeta(3)-933120\zeta(5) +2091096)C_F^2-(27648\zeta(3)+240448)T_F^2N_f^2 ]\frac{C_F\alpha^3}{7776},\\
C_{A,V}&=&1+\mathcal{O}(\alpha^4),\\
C_{44} &=& 1 + 31\frac{C_F \alpha}{9} + \left[ \left( - 1782 \zeta(3) + 6404 \right) C_A + \left( 1296 \zeta(3) - 228 \right)
C_F - 2668 T_F N_f \right] \frac{C_F \alpha^2}{162} \nonumber \\
&& + \left[ \left(  
- 11944044 \zeta(3) + 746496 \zeta(4) +524880 \zeta(5) + 38226589 \right) C_A^2 \right. 
\nonumber \\
&& \left.+ \left(  - 4914432 \zeta(3) - 2239488 \zeta(4) + 8864640 \zeta(5) 
+ 3993332 \right) C_A C_F \right. \nonumber \\
&& \left. + \left(  369792 \zeta(3) - 1492992 \zeta(4) - 24752896 \right) C_A T_F N_f 
\right. \nonumber \\
&& \left. + \left( 10737792 \zeta(3) 
+ 1492992 \zeta(4) - 9331200 \zeta(5) - 3848760 \right) C_F^2 \right. 
\nonumber \\
&& \left. - \left( - 3234816 \zeta(3) - 1492992 \zeta(4) + 9980032 \right) C_F T_F N_f 
\right. \nonumber \\
&& \left. + \left( 221184 \zeta(3) + 3391744 \right) T_F^2 N_f^2 \right] 
\frac{C_F \alpha^3}{69984} + O(a^4),
\end{eqnarray}

where $\zeta(n)$ is the Riemann zeta function and for the SU(3) color group, $T_F=\frac{1}{2}$, $C_F=\frac{4}{3}$, $C_A=3$.

Using these expressions to convert our RI'-MOM results at a reference scale of 2 GeV to $\msbar$ values also at 2 GeV leads to the final RCs listed in Table \ref{Table:msbar}.
\bigskip

\begin{center}
\begin{table}[htbp]
\begin{center}
\begin{tabular}{cccccccc}\hline\hline
$\beta$    & $Z_q$     &    $Z_S$ 	      &  $Z_P$  	         & 	$Z_V$  &	    $Z_A$   & $Z_P/Z_S$    & $Z_{44}$ \\\hline\hline
1.90         & 0.761(3)    &  0.723(3)	      &	 0.434(3)          & 0.622(2)	&0.717(1) & 0.600(4)	&  0.973(9)    \\
1.95         & 0.772(2)	& 0.724(4)              &0.462(2) 	          &0.640(2)	&      0.728(2) &0.637(4)	&  0.977(12)  \\
2.10         & 0.789(2)	& 0.727(2)              & 0.523(1)	          &0.687(1)	&	0.757(1) &0.720(4)	 & 1.019(8) \\\hline\hline
\end{tabular}
\end{center}
\caption{Local $N_f=4$ RCs in the $\msbar$ scheme at 2 GeV.\label{Table:msbar}}
\end{table}  
\end{center}


To estimate the effect of the truncation in the perturbative series, we have also converted our results to $\msbar$ at 2 GeV, but starting from
RI'-MOM results at 10 GeV, converting them to $\msbar$ scheme at 10 GeV, and then evolving the $\msbar$ RCs from 10 GeV to 2 GeV using the scale dependence predicted by the renormalization group equation \cite{Gockeler_1999}
\begin{eqnarray}
R_{\mathcal{O}(\mu,\mu_0)}:=\frac{Z_{\mathcal{O}(\mu)}}{Z_{\mathcal{O}(\mu_0)}}=\exp\left\{    -\int^{\bar{g}(\mu^2)}_{\bar{g}(\mu_0^2)}
dg \frac{\gamma(g)}{\beta(g)}\right\}.
\end{eqnarray}

The effect is negligible on $Z_q$ (affecting only the last digit), it is of the order of $3.5\%$ for $Z_S$, $4\%$ for $Z_P$ and $2\%$ for $Z_{44}$. For a
perturbative series, the effect of truncation is relatively small, but compared with the systematic errors, it is far
from being negligible.


\section{Estimation of the systematics}\label{section:systematics} 

The statistical uncertainties affecting the final results are rather small. Typically of the order  of 1\% for $Z_{44}$, and down to $2-5$ per mille for local RCs.
However, the analysis procedure leading to the final values of these RCs is quite complex and involves many non trivial steps and systematic errors 
turn out to be dominant compared to the tiny statistical ones. A very careful study of systematics is thus unavoidable to produce {\it{in fine}} 
reliable and meaningful results. 

Sources of systematic uncertainties are manifold. They arise from the removal of hypercubic corrections, from the running fit, from the chiral extrapolation, and
from the uncertainties on the lattice spacing and on $\Lambda_{QCD}$. We have carefully estimated each source of uncertainties and final results are given
in Table \ref{Table:msbar_syst}. The first parenthesis gives the statistical uncertainty. The second one comes from the systematics due to the hypercubic removal procedure, 
combined with the running fit range. We have varied both the range of $p^2$ used in the hypercubic removal procedure and in the running fit, separately, to estimate the maximal variation on the final RCs value. This leads to the systematics indicated in the second parenthesis. Finally, the last 
number indicates the systematics due to the chiral extrapolation. 
\begin{center}
\begin{table}[htbp]
\begin{center}
\begin{tabular}{c|ccccccc}\hline\hline
$\beta$    & $Z_q$                      &    $Z_S$ 	                         &  $Z_P$  	         & 	$Z_V$              &	    $Z_A$                            & $Z_P/Z_S$    & $Z_{44}$ \\\hline\hline
1.90         & 0.761(3)(5)(3)         &  0.723(3)(5)(9)	      &	 0.434(3)(3)(6)          & 0.622(2)(1)(5)	           &0.717(1)(2)(6)                     & 0.600(4)(4)	(3)&  0.973(9)(7)(30)    \\
1.95         & 0.772(2)(6)(6) 	& 0.724(4)(5)(3)              &0.462(2)(4)(7)  	          &0.640(2)(1)(5)	&      0.728(2)(2)(4)                &0.637(4)(4)(6)	&  0.977(12)(11)(30)  \\
2.10         & 0.789(2)(6)(7) 	& 0.727(2)(5)(4)              & 0.523(1)(4)(1) 	          &0.687(1)	(1)(2)     &	0.757(1)(2)(4)                   &0.720(4)(2)(5)	 & 1.019(8)(6)(30) \\\hline\hline
\end{tabular}
\end{center}
\caption{Final results for $N_f=4$ local RCs in the $\msbar$ scheme at 2 GeV, for each $\beta$ values considered. The first parenthesis gives the statistical uncertainty, the second one the systematics due to the hypercubic removal procedure, combined with the running fit range, and the last number indicates the systematics due to the chiral extrapolation.\label{Table:msbar_syst}}
\end{table}  
\end{center}

The uncertainties on the lattice spacing $a$ and on the value of $\Lambda_{QCD}$ have also been propagated to the RCs and included in the errors quoted in Table \ref{Table:msbar_syst}. We have varied $a$ by $\pm$ 5\% from its central value and  
taken $\Lambda_{QCD}=316(13)$ MeV from Ref.~\cite{Blossier_alphaS_2}. 

Systematics are estimated separately on local and on twist-2 renormalization constants. They indeed behave quite differently, whether it concerns their pion mass dependence,
or their sensitivity to lattice spacing and $\Lambda_{QCD}$. The hypercubic corrections and the running lead to an uncertainty which does not exceed $1\%$.  
The pion mass dependence of all local RCs is weak and the uncertainties associated with the chiral extrapolation 
small. 

The uncertainty on the lattice spacing and on $\Lambda_{QCD}$, propagated to the local RCs, 
gives a very small sensitivity for $Z_q$ and ${Z_P\over Z_S}$, and an effect of 
about $2\%$ for $Z_S, Z_P$ (see Table \ref{Tab:avar} for example).  In particular, the very effect of the uncertainty 
in the lattice spacing is $\sim$ 0.001 for $Z_q$, $Z_S$ and $Z_P$, and can be neglected for $Z_V$, $Z_A$ or $Z_P/Z_S$ 
that do not run with momenta in perturbation.

The situation is a bit different for $Z_{44}$. Uncertainties due to the $H(4)$ corrections and the running fit are of the order of the statistical errors, the dominant source of uncertainties comes clearly from the chiral extrapolation, which induces errors of the order of $3\%$. In addition, the errors on $a$ and $\Lambda_{QCD}$ produce an additional uncertainty on $Z_{44}$ 
of the order of $3\%$.

Finally we have compared our results with the values for local RCs given in Ref.~\cite{Italiani}. These latter have been obtained using the "democratic"
selection of momenta. Restricting our fitting interval to the one used in this reference ("method M1") we find close results, a precise comparison being
however difficult since only statistical errors are reported in ~\cite{Italiani}. 
In addition, taking into account statistical and systematic errors, our results are also compatible with those from \cite{Martha_2013} for $Z_{44}$.

\section{Conclusion}\label{section:conclusion} 

We have presented an original analysis of quark propagator, vertex functions and twist-2 operators renormalization constants
for $N_f=4$ twisted mass fermions. We have implemented a systematic and rigorous procedure to correct for hypercubic lattice artifacts. This non-perturbative method,
avoiding the selection of momenta usually done in this kind of analysis and the use of perturbative formulae, allows to take advantage of all the data and to 
check the running over a wide range of momenta. We have applied our analysis procedure not only to local operators, but also to the
twist-2 operator $O_{44}$. $\mathcal{O}(a^2)$ lattice artifacts have also been efficiently subtracted. 
In order to compare with experimental values obtained for the corresponding matrix elements, all our results, obtained in the RI'-MOM scheme, 
have been converted to the  $\msbar$ scheme at 2 GeV. 
A precise estimate of systematic errors have also been performed and these latter are shown to be dominant in the case of twist-2 operator $O_{44}$. 
Concerns could be raised because of the fact that the RI-MOM scheme requires gauge fixing. There could in principle be fluctuations arising from the Gribov ambiguity. However, several studies explored this idea in the 90s \cite{g1,g2,g3,g4} as well as later on with Ginsparg-Wilson fermions \cite{Gattringer} and have shown that this effect is less than 1\% and thus the dependence on gauge fixing is negligible. Of course one has to mention that all this previous work was in the quenched approximation. 

The method developed here will be applied in the near future to the new gauge configurations, at the physical pion mass, generated by the ETM Collaboration.

\section{Acknowledgments}

We thank P. Boucaud, F. de Soto, X. Du, Z. Liu, K. Petrov and particularly A. Vladikas for useful discussions.
We are also grateful to N. Carrasco, P. Dimopoulos, R. Frezzotti, V. Lubicz and S. Simula for their careful reading of the manuscript and their remarks.
This work was granted access to the HPC resources of CINES and IDRIS under the allocations 2013-052271 and 2014-052271 made by GENCI. Propagator computations have also extensively used CINECA GPUs in the framework of the DECI-9 project DEC09-NPR-LQCD. Most of the analysis has been done 
in Lyon-CCIN2P3, thanks to the massive storage and large CPUs resources provided. We are grateful to the staff members of all these Computing Centers for their constant help. 
This work was supported by the CNRS and the Alexander von Humboldt Foundation (S.Z.).

\end{document}